\documentstyle[amsmath,amssymb,graphicx,wrapfig]{article}

\def\be{\begin{eqnarray}}
\def\ee{\end{eqnarray}}

\def\p{\partial}

\hoffset= - 1.0in         % switch off for draft style
\begin{document}

\hfill ITEP/TH-18/10

\bigskip

\centerline{\Large{The matrix model version of AGT conjecture
and CIV-DV prepotential
}}

\bigskip

\centerline{\it A.Morozov \footnote{ITEP, Moscow, Russia; morozov@itep.ru} and Sh.Shakirov \footnote{ITEP, Moscow, Russia and MIPT, Dolgoprudny, Russia; shakirov@itep.ru} }

\bigskip

\centerline{ABSTRACT}

\bigskip

{\footnotesize
Recently exact formulas were provided for partition function
of conformal (multi-Penner) $\beta$-ensemble
in the Dijkgraaf-Vafa phase, which,
if interpreted as Dotsenko-Fateev correlator of screenings and
analytically continued in the number of screening insertions,
represents generic Virasoro conformal blocks.
Actually these formulas describe the lowest terms of the $q_a$-expansion,
where $q_a$ parameterize the shape of the Penner potential, and are exact in the filling numbers $N_a$.
At the same time, the older theory of CIV-DV prepotential, straightforwardly extended to arbitrary $\beta$
and to non-polynomial potentials, provides an alternative expansion: in powers of $N_a$ and exact in $q_a$. We check that the two expansions coincide in the overlapping region, i.e. for the lowest terms of expansions in both $q_a$ and $N_a$. This coincidence is somewhat non-trivial, since the two methods use different integration contours: integrals in one case are of the B-function (Euler-Selberg) type, while in the other case they are Gaussian integrals. }

\tableofcontents

\section{Introduction}

This paper is a part of extended study of AGT conjecture \cite{AGT} - \cite{MaMoAGT3}, which unifies a number of important subjects \cite{CFT} - \cite{Nolinal}, that were in the focus of theoretical investigations during the last decade. This particular paper is devoted to the recently formulated explicit relation \cite{MaMoAGT1, MaMoAGT2, MaMoAGT3} between the integrated free-field correlators in Virasoro conformal field theory (CFT) on the Riemann sphere
$$ Z_{DF} = \int\limits_{C_1} d z_1 \ldots \int\limits_{C_N} d z_N \ \left< \prod\limits_{a} :e^{\tilde\alpha_{a} \phi(x_a)}: \prod\limits_{i} :e^{b\phi(z_i)}: \right> = $$
\begin{align}
= \prod_{a<b} (x_b-x_a)^{2 \tilde\alpha_{a} \tilde\alpha_{b}} \ \int\limits_{C_1} d z_1 \ldots \int\limits_{C_N} d z_N \prod\limits_{i < j} (z_j - z_i)^{2b^2} \prod\limits_{i,a} (z_i - x_a)^{2 b \alpha_a}
\label{DF}
\end{align}
and conformal blocks -- holomorphic and anti-holomorphic parts of generic (not necessarily free-field) CFT correlators. Actually, this relation was long expected: since the old papers \cite{DF}, there was a belief that any spherical Virasoro conformal block \cite{CFT} can be obtained by appropriate choice of integration contours $C_1, \ldots, C_N$ in the free-field Dotsenko-Fateev integral $Z_{DF}$. However, precise formulation (including the choice of contours) remained a mystery for some time and was made explicit only recently \cite{MaMoAGT2}.

The aim of the present paper is to investigate the partition function $Z_{DF}$ by the methods of matrix models. As noticed long ago \cite{KMMMP}, integrated free-field correlators like (\ref{DF}) can be viewed as natural objects in matrix model theory, which studies ensembles of "eigenvalue" variables $z_1, \ldots, z_N$ with a certain power of Vandermonde determinant in their integration measures
\begin{align}
Z_{DF} \sim \int\limits_{C_1} d z_1 \ldots \int\limits_{C_N} d z_N \prod\limits_{i < j} (z_j - z_i)^{2\beta} \ \exp\left( - \sum\limits_{i = 1}^{N} V(z_i) \right), \ \ \ \ \ \beta = b^2
\end{align}
where in the case of $Z_{DF}$ potential $V(z)$ is of the multi-Penner shape:
\begin{align}
V(z) = - \sum\limits_{a = 1}^{r+1} \alpha_a \log \big( z - x_a \big), \ \ \ \ \ \alpha_a = 2 b \tilde\alpha_a
\end{align}
Models with such a $\beta$-dependent eigenvalue measure are known in literature as $\beta$-ensembles \cite{BetaEnsembles}. For generic values of $\beta$ they are not, strictly speaking, matrix models, since they cannot be induced by integration over any kind of matrices. However, it appears that most of conventional matrix-model methods and technical tools (such as Ward identities and genus expansion) are perfectly applicable for arbitrary values of $\beta$.

Because of this, Dotsenko-Fateev partition functions can be viewed as (and treated essentially by the same methods as) non-Gaussian eigenvalue matrix models \cite{DV,AMM}, only taking into account properly the $\beta$-deformed integration measure. In this paper we consider a pair of alternative approaches to such non-Gaussian integrals: \textbf{1)} based on the quasiclassical expansion and \textbf{2)} based on exact contour integration. Our ultimate aim is to compare the results of these two approaches with each other and elucidate all possible subtleties.

\paragraph{1)} The first method we use in this paper is \textbf{quasiclassical approach} to matrix models: namely, the theory of Dijkgraaf-Vafa (DV) phases \cite{DV,AMM}. This approach exploits the fact, that non-Gaussian integrals
\begin{align}
\int d z_1 \ldots \int d z_N \prod\limits_{i < j} (z_j - z_i)^{2\beta} \ \exp\left( - \dfrac{1}{\hbar} V(z_1, \ldots, z_N) \right)
\label{nongaus}
\end{align}
in the limit $\hbar \rightarrow 0$ are dominated by fluctuations of integration variables near one of the saddle points $(Z_1, \ldots, Z_N)$ -- solutions of the saddle point equation (or, what is the same, equation of motion):
\begin{align}
\dfrac{ \partial V (Z_1, \ldots, Z_N)}{\partial Z_i} = 0
\end{align}
Since for the eigenvalue matrix models the full multivariate potential has a form of a sum $V(z_1, \ldots, z_N) = V(z_1) + \ldots + V(z_N)$, each of components $Z_i$ has to be a critical point of the single-variable potential $V(z)$. Accordingly, the $N$ components of the saddle point $Z_1, \ldots, Z_N$ get divided into $r$ groups:
\begin{align}
\big( Z_1, \ldots, Z_N \big) = \big( \underbrace{\mu_1, \ldots, \mu_1}_{N_1}, \underbrace{\mu_2, \ldots, \mu_2}_{N_2}, \ldots, \underbrace{\mu_{r}, \ldots, \mu_r}_{N_r} \big)
\end{align}
where $\mu_1, \ldots, \mu_r$ denote the critical points of $V(z)$ (i.e, solutions of $V^{\prime}(\mu_a) = 0$) and $r$ denotes their total number. The parameters $N_a$ satisfy $N_1 + \ldots + N_r = N$ and label the choice of physical phase for the multi-Penner $\beta$-ensemble. After the choice of particular set $\{N_1, \ldots, N_r\}$, one needs to integrate over all fluctiations
\be
\big( z_1, \ldots, z_N \big) = \big( Z_1, \ldots, Z_N \big) + \big( y_1, \ldots, y_N \big)
\ee
treating the corresponding integrals over fluctuation variables $ y_1, \ldots, y_N $ as Gaussian correlators. This method is quite general and applicable to generic potentials; applying this method to particular non-Gaussian integral (\ref{DF}), one obtains an answer for (\ref{DF}) in terms of power series in the small parameter $\hbar$.

\paragraph{2)} The second method is to \textbf{specify concrete integration contours} for the non-Gaussian integral and integrate over these contours directly. Sometimes called theory of integral discriminants \cite{Nolinal}, this approach is rarely used in the literature: since the methods of exact integration for non-Gaussian potentials are under-developed, their application in practice is feasible only for simple enough potentials $V(z)$. Fortunately, the potential $V(z)$ we are interested in -- the multi-Penner potential -- has a very special form: it is a logarithm of a polynomial (for all natural parameters $\alpha_a$). Consequently, the exponent $\exp\big( - V(z)\big)$ is merely a polynomial, so that exact non-Gaussian integration for the partition function (\ref{DF}) is feasible in practice.

At the first glance, the choice of integration contours $C_1, \ldots, C_N$ seems unrestricted, since it is quite easy to integrate a polynomial along any contour in the complex plane. However, as usual in contour integration, homotopic choices of contours actually give one and the same answer for the integral. Because of this, the set of essentially inequivalent choices consists of all segments $\gamma_{ab} = [x_a, x_b]$, which connect the roots of the integrand. Moreover, the obvious relations $\gamma_{ab} + \gamma_{bc} = \gamma_{ac}$ leave only $r$ inequivalent contours $\gamma_a = [x_1, x_a]$ at our disposal, so that the $N$ contours $C_1, \ldots, C_N$ get divided into $r$ groups:

\begin{align}
\big( C_1, \ldots, C_N \big) = \big( \underbrace{\gamma_2, \ldots, \gamma_2}_{N_1}, \underbrace{\gamma_3, \ldots, \gamma_3}_{N_2}, \ldots, \underbrace{\gamma_{r+1}, \ldots, \gamma_{r+1}}_{N_r} \big)
\end{align}
\smallskip\\
Note, that for any natural $N_a, \alpha_a$ and $\beta$ the integrand is a polynomial, hence the integral can be evaluated exactly. Note also, that (just like in the first approach) additional parameters $N_a$ appear naturally. They satisfy  $N_1 + \ldots + N_r = N$ and label the choice of phase for the multi-Penner $\beta$-ensemble. One concludes that this appearance is inavoidable, reflects a natural property of non-Gaussian integrals (they are not fully defined by the action alone) and should be reproduced by any other consistent approach to non-Gaussian ensembles.

Applying this method to the non-Gaussian integral (\ref{DF}), one obtains an answer for (\ref{DF}) in terms of power series in the small parameters $q_a$ (related to $x_a$ by a simple law, see \cite{MaMoAGT2}). These series were first evaluated in \cite{MaMoAGT2,MaMoAGT3} and found to coincide with the standard conformal blocks in Virasoro CFT, thus providing an explicit and convincing check of the "matrix-model version" of the AGT conjecture. The observation of \cite{MaMoAGT2} has lately led to a number of developments \cite{MaMoAGT3}, aimed at proving the original AGT conjecture by matrix-model methods. This paper can be considered as yet another contribution to this direction of research.

\paragraph{1) vs. 2)} In the present paper, we adress the issue of the relation between two different methods 1) and 2), and two corresponding expansions -- in powers of $\hbar$ (the quasiclassical DV expansion) and in powers of $q$ (the CFT expansion). Both methods can be applied to calculate the partition function of the multi-Penner $\beta$-ensemble $Z_{DF}$. Despite they are designed to calculate one and the same quantity -- partition function of the non-Gaussian $\beta$-ensemble -- it is not \emph{a priori} obvious that their results are consistent with each other.

Actually, the reason for doubts is severe difference in the choice of integration contours. The first method (selection of contours and exact non-Gaussian integration over these contours) naturally incorporates the choice $\gamma_a = [x_1, x_a]$ of integration contours for the variables $z_1, \ldots, z_N$. Such contours are not quite typical for matrix-model theory: usually, one integrates either along closed contours or along open contours that go to infinity, like the real axis in the Hermitian matrix model, the imaginary axis for the anti-Hermitian model, etc. As explained in section 4, precisely such "typical matrix-model" contours are used in the second method for integration over the fluctuation variables $y_1, \ldots, y_N $. It is not \emph{a priori} clear, why such different treatments of a non-Gaussian integral would give one and the same answer.

Therefore, consistency between the two methods is not quite trivial and deserves to be checked. It is a goal of present paper to perform such a check. To simplify our presentation, we consider only the simplest case of the 3-Penner potential (which corresponds on the CFT side to the 4-point conformal block on a sphere):

\begin{align}
V(z) = - \alpha_1 \log (z) - \alpha_2 \log (z - q) - \alpha_3 \log( z - 1)
\label{3Penner}
\end{align}
\smallskip\\
The first method provides a series expansion in a single small parameter $q$, described in \cite{MaMoAGT2,MaMoAGT3}. The second method provides a series expansion in $\hbar$, derived in this paper. To compare both methods, we study the double $(\hbar,q)$-expansion from both the contour integration and quasiclassics sides. \textbf{We find complete agreement}.

This paper is organized as follows. We begin in section 2 with a simple (perhaps even oversimplified) example of comparison between \textbf{1)} the contour-integration and \textbf{2)} the quasiclassical DV methods. This example illustrates and helps to clarify our claims. Then, in section 3, we describe the DV method for arbitrary non-polynomial potentials in full generality. The particular case of logarithmic 3-Penner potential is considered in section 4: using the DV method, we obtain $\hbar$-expansion of the 3-Penner partition function and derive its double $(\hbar,q)$ expansion. Finally, in section 5, we consider the $q$-expansion of the same 3-Penner partition function and derive the double $(\hbar,q)$ expansion. \emph{A priori} different, the two double $(\hbar,q)$ expansions are found to coincide.

\pagebreak

The Appendix of the present paper is devoted to another important special case of the general considerations of section 3: namely, the case of polynomial potentials
\begin{align}
V(z) = \sum\limits_{a = 1}^{r+1} T_a z^{a}
\end{align}
Models with such potentials recieved a considerable amount of attention \cite{DV} some time ago, because at $\beta = 1$ their free energy $F = \log Z$ makes appearance in supersymmetric 4d gauge theories with spontaneously broken gauge symmetry. The quasiclassical expansion, which we use most often in the present paper, is actually not the most convenient for comparison with 4d gauge theories: more convenient is the so-called genus expansion, that is, the same quasiclassical expansion expressed in terms of different variables $S_a = \hbar N_a$. In terms of 4d gauge theory, variables $S_a$ posess an interpretation of vaccuum condensates of gauge fields, while variables $N_a$ do not literally correspond to any gauge theory quantities. In terms of variables $S_a$, the free energy takes form

\begin{align}
\beta = 1: \ \ \ F = \log Z = \dfrac{1}{\hbar^2} \Big( \ F_0(S_1, \ldots, S_r) + \hbar^2 F_1(S_1, \ldots, S_r) + \ldots \ \Big)
\label{CIV-DV}
\end{align}
\smallskip\\
where the components $F_p(S_1, \ldots, S_r)$ are called genus-$p$ prepotentials, and the leading genus zero component $F_0(S_1, \ldots, S_r)$ is widely known as CIV-DV prepotential. Using the general formulas of section 3, we calculate the CIV-DV prepotential (for arbitrary degree $r = \deg V(z)$ of the potential) and generalize it in two directions: to higher genera and to arbitrary $\beta$. For $\beta \neq 1$, we observe appearance of contributions of odd order

\begin{align}
\beta \neq 1: \ \ \ F = \log Z = \dfrac{1}{\hbar^2} \Big( \ F_0(S_1, \ldots, S_r) + \hbar F_{1/2}(S_1, \ldots, S_r) + \hbar^2 F_1(S_1, \ldots, S_r) + \ldots \ \Big)
\label{BETA-CIV-DV}
\end{align}
\smallskip\\
which correspond to non-integer genera in the genus expansion. We derive explicit expressions for the generalized $\beta$-deformed CIV-DV prepotentials of genera $0$, $1/2$, $1$ and $3/2$, and study some of their simplest properties. Despite these expressions have little in common with the main topic of present paper, they are interesting by theirselves and will be probably useful in future research. We include them for reference reasons.

\section{A toy model: B-function}

To illustrate the difference between the choices of integration contours in methods \textbf{1)} and \textbf{2)}, let us now consider an oversimplified example: the B-function integral
\begin{align}
B(u,v) = \int z^u (1 - z)^v dz
\end{align}
This integral, actually, corresponds to the particular case $\alpha_1 = u, \alpha_2 = 0, \alpha_3 = v$ and $N = 1$ of the 3-Penner integral. From the point of view of the method \textbf{1)}, as explained above, we need to choose essentially independent integration contours. We can choose one such contour -- connecting the points 0 and 1. With this choice of contour, the integral becomes correctly defined and convergent, giving an exact answer
\begin{align}
B_1(u,v) = \int\limits_{0}^{1} z^u (1 - z)^v dz = \dfrac{\Gamma(u + 1)\Gamma(v + 1)}{\Gamma(u + v + 1)}
\end{align}
As already mentioned above, this contour is not typical for matrix models; a typical matrix-model contour would be $(-\infty, \infty)$. From the point of view of the method \textbf{2)}, we need to find the critical points of
\begin{align}
V(z) = - u \log z - v \log( 1 - z)
\end{align}
i.e, solutions of the equation of motion
\begin{align}
V^{\prime}(z) = - \dfrac{u}{z} - \dfrac{v}{1 - z} = 0
\end{align}
It is easy to see that the only solution (the only critical point) is $u/(u + v)$. To describe quasiclassical fluctuations of the single variable $z$ around this critical point, we need to make a change of variable
\begin{align}
z = \dfrac{u}{u + v} + y
\end{align}
where $y$ describes the fluctuation. The DV procedure, just like any other quasiclassical method, prescribes to integrate the fluctuation from $-\infty$ to $+\infty$. Accordingly, the integral takes form (we omit $\hbar$ for simplicity)
\begin{align}
B_2\left(u, v\right) = \dfrac{u^u v^v}{(u+v)^{u+v}} \int\limits_{-\infty}^{\infty} \exp\left( - \dfrac{(u+v)^3}{2uv} y^2 - \dfrac{(u+v)^4(u-v)}{3u^2v^2} y^3 + \ldots \right) dy
\end{align}
The point is that \emph{a priori} $B_1(u,v)$ and $B_2(u,v)$ are different integrals, since the contours of integration differ. However, they do coincide as expansions in $u,v >> 1$ (equivalently, as expansions in $\hbar \rightarrow 0$). One can check this directly by evaluating $B_2\left(u, v\right)$ as a Gaussian integral
\begin{align}
B_2\left(u, v\right) = \dfrac{u^u v^v}{(u+v)^{u+v}} \times \sqrt{ \dfrac{2uv}{(u+v)^3} } \times \sqrt{2\pi} \times \left( 1 + \dfrac{v^2 - 11 uv + u^2}{12uv(u+v)} + \ldots \right)
\end{align}
what reproduces the well-known Stirling expansion of $B_1(u,v)$ at $u,v >> 1$. In fact, it is hard to expect a different result in this toy model: we consider just a quasiclassical expansion of a single-variable integral. What we check in present paper, is that the coincidence holds further for $N > 1$ and $\alpha_2 \neq 0$.

\paragraph{} We now mention another way to derive the relation between Gaussian and B-function (hypergeometric) integrals, which may be more straightforward but somewhat delicate as soon as one touches the issues of integration contours
and analytical continuation. From the definition of the exponent we have

\be
\int_{-\infty}^\infty z^{2p} e^{-z^2} dz =
\lim_{M\rightarrow \infty}
\int_{-\infty}^\infty z^{2p} \left(1-\frac{z^2}{M}\right)^M dz
= \lim_{M\rightarrow \infty} M^{p+1/2}\int_0^\infty
x^{p-1/2} (1-x)^M dx
\label{gahy}
\ee
\smallskip\\
If the integral at the r.h.s. was between $0$ to $1$,
then the r.h.s. would be

\be
\lim_{M\rightarrow \infty} M^{p+1/2}\frac{\Gamma(M+1)\Gamma(p+1/2)}
{\Gamma(M+p+3/2)} = \Gamma(p+1/2) = \frac{(2p-1)!!\sqrt{\pi}}{2^pp!}
\ee
\smallskip\\
what is the correct answer for the original Gaussian integral.
If integral at the r.h.s. of (\ref{gahy}) was indeed taken
from $0$ to $\infty$, it would diverge, but one could split
it into two, from $0$ to $1$ and from $1$ to $\infty$.
The latter one could then be analytically continued with the
help of modular transformation $x\rightarrow 1/x$:

\be
\int_1^\infty  x^{p-1/2} (1-x)^M dx \rightarrow
(-)^M \int_0^1  x^{-M-p-3/2} (1-x)^M dx = \frac{(-)^M\Gamma(-M-p-1/2)\Gamma(M+1)}{\Gamma(1/2-p)} =
\ee
\be
= \frac{(-)^M\sin(p+1/2)\pi}{\sin(M+p+3/2)\pi}
\frac{\Gamma(M+1)\Gamma(p+1/2)}{\Gamma(M+p+3/2)} =
- \int_0^1  x^{p-1/2} (1-x)^M dx
\ee
\smallskip\\
so that adding the two pieces one would get a vanishing answer.
Of course, {\bf the right answer is to declare that the
hypergeometric integral} at the r.h.s. of (\ref{gahy})
is a $B$-function, i.e. {\bf goes from $0$ to $1$ only
-- and this is
exactly what we observe in this paper in more general situation.}
In the simple context of (\ref{gahy}) one can easily justify
this prescription. Indeed, let original Gaussian integral be
taken between $-K$ and $K$ with $K\gg 1$ but finite: this is
a good approximation, of course. Then the $x$-integral in
(\ref{gahy}) will go between $0$ and $\sqrt{K/M}$, and this upper limit tends to
zero if $M\rightarrow \infty$ with $K$ fixed. Thus there is
definitely no need to include the region $1\leq x <\infty$
into the integration domain. On another side it can be continued
from $\sqrt{K/M}\ll 1$ to $1$, because for large $M$ only
the narrow region near $x=0$ actually contributes.

\pagebreak

\section{$\hbar$-expansion for the generic case}

For arbitrary potential $V(z)$, one can define and study the non-Gaussian partition function

\begin{align}
Z = \int\limits d z_1 \ldots \int\limits d z_N \prod\limits_{i < j} (z_j - z_i)^{2\beta} \ \exp\left( - \dfrac{1}{\hbar} \sum\limits_{i = 1}^{N} V(z_i) \right)
\label{PartitionFunction}
\end{align}
\smallskip\\
Such models have been studied for quite a long time: in the case, when $V(z)$ is a polynomial, their free energy $F = \log Z$ gives rise to the CIV-DV prepotential \cite{DV}. Throughout this section, we do \emph{not} assume that $V(z)$ is polynomial; we assume only that $V(z)$ admits an infinite Taylor series expansion near each of its critical points. In particular, all polynomials and all logarithmic (multi-Penner) potentials do fall into this class, while potentials of fractional degree (like $(z - \mu)^{3/2}$) do not.

It is most important to note that, since the integral is non-Gaussian, it is not completely defined by eq. (\ref{PartitionFunction}): in addition, it depends on the choice of integration contours. There are at least two different ways to remove this ambiguity: either to select a particular contour (actually, one can associate an independent contour $C_i$ with every variable $z_i$) or to study the integral quasiclassically around some particular saddle point. This section is devoted to the second -- quasiclassical -- option.

The saddle point analysis of the integral (\ref{PartitionFunction}) is quite simple: when $\hbar \rightarrow 0$, variables $z_i$ become distributed between the critical points -- solutions of the equation $V^{\prime}(z) = 0$. For arbitrary $V(z)$, this equation can have infinitely many solutions. Two classes of potentials, often encountered in practice, posess a finite number $r$ of critical points $\mu_1, \ldots, \mu_r$: polynomial potentials
\be
V(z) = \sum\limits_{a = 1}^{r+1} T_a z^a
\ee
and logarithmic (multi-Penner) potentials
\begin{align}
V(z) = - \sum\limits_{a = 1}^{r+1} \alpha_a \log \big( z - x_a \big)
\end{align}
It is easy to see that different saddle points of the non-Gaussian integral are labeled by different partitions $N = N_1 + \ldots + N_r$. Indeed, the only possibility modulo permutations is that the first $N_1$ eigenvalues tend to the critical point $\mu_1$, then next $N_2$ eigenvalues tend to $\mu_2$, and so on:

\be
(Z_1, \ldots, Z_N) = \big(\underbrace{\mu_1, \ldots, \mu_1}_{N_1}, \ldots, \underbrace{\mu_r, \ldots, \mu_r}_{N_r}\big)
\ee
\smallskip\\
as illustrated on Fig.1. The numbers $N_i$ are called filling numbers. To evaluate the integral quasiclassically, we need to study the integrand in the vicinity of the saddle point:

\begin{figure}[t]
  \begin{center}
\includegraphics[width=0.39\textwidth]{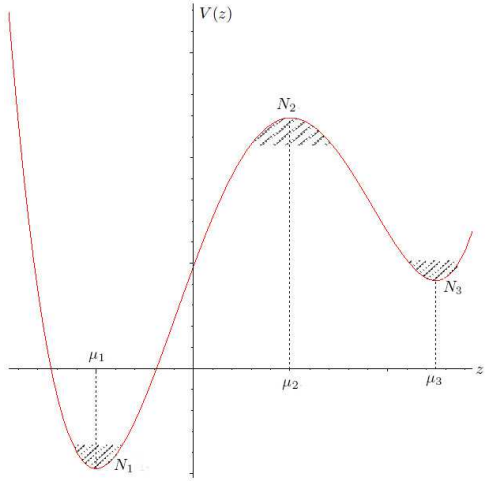}
  \end{center}
  \caption{Filling numbers $N_i$ describe distribution of eigenvalues between critical points $\mu_i$ of $V(z)$.}
  \bigskip
\end{figure}
\be
(z_1, \ldots, z_N) = \big(\underbrace{\mu_1, \ldots, \mu_1}_{N_1}, \ldots, \underbrace{\mu_r, \ldots, \mu_r}_{N_r}\big) + \big(\underbrace{y^{(1)}_1, \ldots, y^{(N_1)}_1}_{N_1}, \ldots, \underbrace{y^{(1)}_r, \ldots, y^{(N_r)}_r}_{N_r}\big)
\ee
\smallskip\\
Formally, this can be seen as a change of integration variables, where the $y$-variables describe fluctuations near the saddle point. Doing this change of integration variables, we obtain the following integral:

$$ Z = \prod\limits_{a = 1}^{r} \prod\limits_{i = 1}^{N_a} \int d y^{(i)}_a \exp\left( - \dfrac{1}{\hbar} V\big(\mu_a + y^{(i)}_a\big) \right) \times \emph{}
$$
\be
\emph{} \times \prod\limits_{a = 1}^{r} \prod\limits_{1 \leq i < j \leq N_a} \left(y^{(i)}_a - y^{(j)}_a\right)^{2\beta} \cdot \prod\limits_{1 \leq a < b \leq r} \prod\limits_{i = 1}^{N_a} \prod\limits_{j = 1}^{N_b} \left(\mu_a - \mu_b + y^{(i)}_{a} - y^{(j)}_b\right)^{2\beta}
\ee
\smallskip\\
Expanding the potential into Taylor series, we find
$$ Z = \prod\limits_{a = 1}^{r} \prod\limits_{i = 1}^{N_a} \int d y^{(i)}_a \exp\left( - \dfrac{1}{\hbar} V\big(\mu_a) - \dfrac{1}{\hbar} \sum\limits_{k = 2}^{\infty} \dfrac{\big(y^{(i)}_a\big)^k}{k!} V^{(k)}\big(\mu_a) \right) \times \emph{}
$$
\be
\emph{} \times \prod\limits_{a = 1}^{r} \prod\limits_{1 \leq i < j \leq N_a} \left(y^{(i)}_a - y^{(j)}_a\right)^{2\beta} \cdot \prod\limits_{1 \leq a < b \leq r} \prod\limits_{i = 1}^{N_a} \prod\limits_{j = 1}^{N_b} \left(\mu_a - \mu_b + y^{(i)}_{a} - y^{(j)}_b\right)^{2\beta}
\ee
This integral can be written as
$$ Z = {\rm const} \times \prod\limits_{a = 1}^{r} \prod\limits_{i = 1}^{N_a} \int d y^{(i)}_a \exp\left( - \dfrac{1}{\hbar} \sum\limits_{k = 2}^{\infty} \dfrac{\big(y^{(i)}_a\big)^k}{k!} V^{(k)}\big(\mu_a) \right) \times \emph{}
$$
\be
\emph{} \times \prod\limits_{a = 1}^{r} \prod\limits_{1 \leq i < j \leq N_a} \left(y^{(i)}_a - y^{(j)}_a\right)^{2\beta} \cdot \prod\limits_{1 \leq a < b \leq r} \prod\limits_{i = 1}^{N_a} \prod\limits_{j = 1}^{N_b} \left(1 + \dfrac{y^{(i)}_{a} - y^{(j)}_b}{\mu_a - \mu_b} \right)^{2\beta}
\ee
where the normalisation constant is defined as
\be
{\rm const} = \prod\limits_{a = 1}^{r} \exp\left( - \dfrac{N_a}{\hbar} V\big(\mu_a) \right) \cdot \prod\limits_{1 \leq a < b \leq r} (\mu_a - \mu_b)^{2\beta N_a N_b}
\ee
The variables $y^{(1)}_a \ldots y^{(N_a)}_a$ can be considered as eigenvalues of $N_a \times N_a$ matrices $\Phi_a$, where $a$ takes values $ 1,\ldots, r$. In terms of $\Phi_a$, the integral is slightly simplified: introducing the $\beta$-deformed Vandermonde measure
\be
D_{\beta}\Phi_a = \prod\limits_{1 \leq i < j \leq N_a} \left(y^{(i)}_a - y^{(j)}_a\right)^{2\beta} [dU] \prod\limits_{i = 1}^{N_a} \int d y^{(i)}_a, \ \ \ \mbox{ where} \ \ \Phi_a = U \ {\rm diag}(y^{(1)}_a \ldots y^{(N_a)}_a) \ U^{+}
\ee
and introducing a convenient trace-like notation for the symmetric combinations of variables
\be
\sum\limits_{i = 1}^{N_a} \big(y^{(i)}_a\big)^k = {\rm Tr} \ (\Phi_a)^k
\ee
we can represent the measure (including its potential part) in these new matrix variables:
$$
 \prod\limits_{a = 1}^{r} \prod\limits_{i = 1}^{N_a} \int d y^{(i)}_a \exp\left( - \dfrac{1}{\hbar} \sum\limits_{k = 2}^{\infty} \dfrac{\big(y^{(i)}_a\big)^k}{k!} V^{(k)}\big(\mu_a) \right) \prod\limits_{a = 1}^{r} \prod\limits_{1 \leq i < j \leq N_a} \left(y^{(i)}_a - y^{(j)}_a\right)^{2\beta} = \emph{}
$$
\be
\emph{} = \prod\limits_{a = 1}^{r} \int D_{\beta}\Phi_a \exp\left( - \dfrac{1}{\hbar} \sum\limits_{k = 2}^{\infty} \dfrac{{\rm Tr} \ (\Phi_a)^k}{k!} V^{(k)}\big(\mu_a) \right)
\ee
The second factor, which mixes different matrices with each other, can be also represented in terms of traces. To do this, one needs to represent this factor as exponent of a logarithm
$$
\prod\limits_{1 \leq a < b \leq r} \prod\limits_{i = 1}^{N_a} \prod\limits_{j = 1}^{N_b} \left(1 + \dfrac{y^{(i)}_{a} - y^{(j)}_b}{\mu_a - \mu_b} \right)^{2\beta} = \exp\left( 2 \beta \sum\limits_{1 \leq a < b \leq r} \sum\limits_{i = 1}^{N_a} \sum\limits_{j = 1}^{N_b} \log\left(1 + \dfrac{y^{(i)}_{a} - y^{(j)}_b}{\mu_a - \mu_b} \right) \right)
$$
then formally expand the resulting logaritm into Taylor series
$$
\exp\left( 2 \beta \sum\limits_{1 \leq a < b \leq r} \sum\limits_{i = 1}^{N_a} \sum\limits_{j = 1}^{N_b} \log\left(1 + \dfrac{y^{(i)}_{a} - y^{(j)}_b}{\mu_a - \mu_b} \right) \right) = \exp\left( - 2 \beta \sum\limits_{1 \leq a < b \leq r} \sum\limits_{i = 1}^{N_a} \sum\limits_{j = 1}^{N_b} \sum\limits_{k = 1}^{\infty} \dfrac{(-1)^k}{k} \left( \dfrac{y^{(i)}_{a} - y^{(j)}_b}{\mu_a - \mu_b} \right)^k \ \right)
$$
and finally convert the symmetric combinations of variables $y^{(i)}_{a}$ into traces of matrix variables $\Phi_a$:
$$
\exp\left( - 2 \beta \sum\limits_{1 \leq a < b \leq r} \sum\limits_{i = 1}^{N_a} \sum\limits_{j = 1}^{N_b} \sum\limits_{k = 1}^{\infty} \dfrac{(-1)^k}{k} \left( \dfrac{y^{(i)}_{a} - y^{(j)}_b}{\mu_a - \mu_b} \right)^k \ \right) = \emph{}
$$
$$
\emph{} = \exp\left( - 2 \beta \sum\limits_{1 \leq a < b \leq r} \sum\limits_{p + q > 0} \dfrac{(-1)^{p}(p+q-1)!}{p!q!} \dfrac{{\rm Tr} \ (\Phi_a)^p \ {\rm Tr} \ (\Phi_b)^q}{(\mu_a - \mu_b)^{p+q}} \right)
$$
After these algebraic transformations, the partition function is rewritten purely in terms of matrix variables:
$$
Z = {\rm const} \times \int D_{\beta}\Phi_1 \ldots \int D_{\beta}\Phi_r \ \exp\left( - \dfrac{1}{\hbar} \sum\limits_{a = 1}^{r} \sum\limits_{k = 2}^{\infty}  V^{(k)}\big(\mu_a) \dfrac{{\rm Tr} \ (\Phi_a)^k}{k!} - \emph{} \right.
$$
\be
\left. \emph{} - 2 \beta \sum\limits_{1 \leq a < b \leq r} \sum\limits_{p + q > 0} \dfrac{(-1)^{p}(p+q-1)!}{p!q!} \dfrac{{\rm Tr} \ (\Phi_a)^p \ {\rm Tr} \ (\Phi_b)^q}{(\mu_a - \mu_b)^{p+q}} \right)
\ee
It is also convenient to rescale these matrix variables in the following way:
\be
\Phi_a = \phi_a \sqrt{ \dfrac{\hbar}{\Delta_a} }, \ \ \ \ {\rm where} \ \Delta_a = V^{(2)}\big(\mu_a) = V^{\prime \prime}(\mu_a)
\ee
Such a rescaling makes the coefficients in front of the Gaussian terms ${\rm Tr} \ (\Phi_a)^2$ equal to unity, what is convenient for the subsequent perturbative expansion. Ultimately, we obtain an $r$-matrix Gaussian integral

\begin{align}
\boxed{ \ \ \
Z = Z_0 \times \int D_{\beta}\phi_1 \ldots \int D_{\beta}\phi_r \ \exp\left( - \dfrac{{\rm Tr} \ (\phi_1)^2}{2} - \ldots - \dfrac{{\rm Tr} \ (\phi_r)^2}{2} - W\big( \phi_1, \ldots, \phi_r \big) \right) \ \ \ }
\end{align}
\smallskip\\
with peculiar interaction
\be
W\big( \phi_1, \ldots, \phi_r \big) = \sum\limits_{a = 1}^{r} \sum\limits_{k = 3}^{\infty} A_k(\mu_a) {\rm Tr} \ (\phi_a)^k + \sum\limits_{1 \leq a < b \leq r} \sum\limits_{p + q > 0} B_{pq}(\mu_a, \mu_b) {\rm Tr} \ (\phi_a)^p \ {\rm Tr} \ (\phi_b)^q
\label{Interaction}
\ee
where
\be
A_k(\mu_a) = \dfrac{1}{\hbar} \dfrac{V^{(k)}\big(\mu_a)}{k!} \left( \dfrac{\hbar}{\Delta_a} \right)^{k/2}
\ee
\be
B_{pq}(\mu_a, \mu_b) = 2 \beta \ \dfrac{(-1)^{p}(p+q-1)!}{p!q!} \dfrac{1}{(\mu_a - \mu_b)^{p+q}} \left( \dfrac{\hbar}{\Delta_a} \right)^{p/2} \left( \dfrac{\hbar}{\Delta_b} \right)^{q/2}
\ee
and normalisation constant
\be
Z_0 = \prod\limits_{a = 1}^{r} \exp\left( - \dfrac{N_a}{\hbar} V\big(\mu_a) \right) \left( \dfrac{\hbar}{\Delta_a} \right)^{\dfrac{\beta N_a(N_a - 1) + N_a}{2}} \cdot \prod\limits_{1 \leq a < b \leq r} (\mu_a - \mu_b)^{2\beta N_a N_b}
\label{NormConst}
\ee
One can see, that at $\hbar \rightarrow 0$ the potential $W$ is suppressed by powers of $\hbar$. Thus, in accordance with general quasiclassical prescriptions, the partition function $Z$ can be treated perturbatively, as a Gaussian integral

\begin{align}
Z = Z_0 \times \int D_{\beta}\phi_1 \ldots \int D_{\beta}\phi_r \ \left( 1 + W + \dfrac{W^2}{2} + \dfrac{W^3}{6} + \ldots \right) \ \exp\left( - \dfrac{{\rm Tr} \ (\phi_1)^2}{2} - \ldots - \dfrac{{\rm Tr} \ (\phi_r)^2}{2} \right)
\label{WPerturb}
\end{align}
\smallskip\\
Substituting the expression (\ref{Interaction}) into the expansion (\ref{WPerturb}) and gathering terms of equal $\hbar$-degree, we find
\be
\boxed{ \ \
Z = Z_0 \times \prod\limits_{a = 1}^{r} {\rm Vol}_\beta(N_a) \times \left( 1 + \sum\limits_{k > 0} H_k \hbar^k \right)
\ \ }
\label{DV1}
\ee
This formula is, basically, the main output of the DV method: evaluation of the multiple non-Gaussian integral (\ref{PartitionFunction}) in terms of power series in $\hbar$.
It has three consistutients. The first is normalisation constant $Z_0$, which represents the semiclassical contribution (the value of the integrand on the selected saddle point, i.e, on the selected solution $(Z_1, \ldots, Z_N)$ of equations of motion). The second is a product of $\beta$-deformed volumes
\be
{\rm Vol}_\beta(N) = \int\limits_{N \times N} D_{\beta}\phi \ \exp\left( - \dfrac{{\rm Tr} \ (\phi)^2}{2} \right) = (\sqrt{2 \pi})^N \prod\limits_{k = 1}^{N} \dfrac{\Gamma(1 + \beta k)}{\Gamma(1 + \beta)}
\label{DV2}
\ee
of the unitary groups $U(N_1), \ldots, U(N_r)$. This product is the contribution of the terms of order $W^0$ in (\ref{WPerturb}), and all the higher contributions are proportional to this product. It is necessary to note, that ${\rm Vol}_\beta(N)$ has the meaning of the volume of the unitary group only for $\beta = 1$; for generic $\beta$, we are not aware of any simple group-theoretical representation. Finally, the third contitutient is the quantum series
\begin{align*}
Z_{\hbar} = 1 + \sum\limits_{k > 0} H_k \hbar^k
\end{align*}
which represent perturbative in $\hbar$ quantum corrections to the leading (nonperturbative in $\hbar$) classical part $Z_0$. Each particular coefficient $H_k$ can be straightforwardly calculated: for example, we find

\begin{align}
\nonumber H_1 \ = \ & \dfrac{1}{72} \sum\limits_{a} \dfrac{V^{(3)}\big(\mu_a)C_{33}(N_a)}{\Delta_a^3} - \dfrac{1}{24} \sum\limits_{a} \dfrac{V^{(4)}\big(\mu_a)C_{4}(N_a)}{\Delta_a^2} + \dfrac{\beta }{3} \sum\limits_{a \neq b} \dfrac{V^{(3)}\big(\mu_a)C_{13}(N_a)C_0(N_b)}{\Delta_a^2 (\mu_a - \mu_b)} + \emph{} & \\ \nonumber & \\ & \emph{} + \sum\limits_{a \neq b} \dfrac{2 \beta^2 C_{11}(N_a)C_{00}(N_b) + \beta C_2(N_a) C_0(N_b)}{\Delta_a (\mu_a - \mu_b)^2} +  \sum\limits_{a \neq b \neq c} \dfrac{2 \beta^2 C_{11}(N_a)C_{0}(N_b)C_0(N_c)}{\Delta_a (\mu_a - \mu_b) (\mu_a - \mu_c)}
\label{DV4}
\end{align}
\smallskip\\
where $C_{k_1, \ldots, k_m}$ are the Gaussian correlators

\begin{align}
C_{k_1, \ldots, k_m}(N) = \dfrac{1}{{\rm Vol}_\beta(N)} \int\limits_{N \times N} D_{\beta}\phi \ {\rm Tr} \ (\phi)^{k_1} \ldots {\rm Tr} \ (\phi)^{k_m} \ \exp\left( - \dfrac{{\rm Tr} \ (\phi)^2}{2} \right)
\label{Correl}
\end{align}
\smallskip\\
with $\beta$-dependent eigenvalue measure. We compute these Gaussian correlators, using the Ward identities for the integral (\ref{Correl}), which can be written in the form of recursive relations
\begin{align}
& C_{k_1, \ldots, k_m} = \beta \sum\limits_{a + b = k_m - 2} C_{k_1, \ldots, k_{m-1}, a, b} + \sum\limits_{l = 1}^{m} C_{k_1, \ldots, k_l + k_m - 2, \ldots, k_{m-1}} + (1 - \beta) (k_m - 1) C_{k_1, \ldots, k_m - 2}
\end{align}
Since each step of induction lowers the total sum of indices $k_1 + \ldots + k_m$ by 2, Ward identities allow to calculate each particular correlator $C_{k_1, \ldots, k_m}(N)$ in no more than $(k_1 + \ldots + k_m)/2$ steps. This gives a convenient and simple computer algorithm to calculate these correlators. Several simplest correlators for even $k_1 + \ldots + k_m$ are

\begin{align*}
{\rm level} \ 2: \ & C_{2} = (1-\beta) N+N^2 \beta, \ \ \ \ C_{1,1} = N \\
\end{align*}
\begin{align*}
{\rm level} \ 4: \ &  C_{4} = (3-5 \beta+3 \beta^2) N+(5 \beta-5 \beta^2) N^2+2 N^3 \beta^2, \ \ \ \ C_{3,1} = (3-3 \beta) N+3 N^2 \beta\\ &
\\ &
C_{2,2} = (2-2 \beta) N+(1+\beta^2) N^2+(2 \beta-2 \beta^2) N^3+N^4 \beta^2\\ &
\\ &
C_{2,1,1} = 2 N+(1-\beta) N^2+N^3 \beta, \ \ \ \ C_{1,1,1,1} = 3N^2 \\
\end{align*}
\begin{align*}
{\rm level} \ 6: \ &  C_{6} = (15-32 \beta+32 \beta^2-15 \beta^3) N+(32 \beta-54 \beta^2+32 \beta^3) N^2+(22 \beta^2-22 \beta^3) N^3+5 N^4 \beta^3\\ &
\\ &
C_{5,1} = (15-25 \beta+15 \beta^2) N+(25 \beta-25 \beta^2) N^2+10 N^3 \beta^2\\ &
\end{align*}
\begin{align*}
& C_{4,2} = (12-20 \beta+12 \beta^2) N+(3+12 \beta-12 \beta^2-3 \beta^3) N^2+(8 \beta-7 \beta^2+8 \beta^3) N^3+(7 \beta^2-7 \beta^3) N^4+2 N^5 \beta^3\\ &
\\ &
C_{3,3} = (15-27 \beta+15 \beta^2) N+(27 \beta-27 \beta^2) N^2+12 N^3 \beta^2\\ &
\\ &
C_{4,1,1} = (12-12 \beta) N+(3+7 \beta+3 \beta^2) N^2+(5 \beta-5 \beta^2) N^3+2 N^4 \beta^2\\ &
\\ &
C_{3,2,1} = (12-12 \beta) N+(3+6 \beta+3 \beta^2) N^2+(6 \beta-6 \beta^2) N^3+3 N^4 \beta^2\\ &
\\ &
C_{3,1,1,1} = 6 N+(9-9 \beta) N^2+9 N^3 \beta \\ &
\\ &
C_{2,2,2} = (8-8 \beta) N+(6-4 \beta+6 \beta^2) N^2+(1+9 \beta-9 \beta^2-\beta^3) N^3+(3 \beta+3 \beta^3) N^4+(3 \beta^2-3 \beta^3) N^5+N^6 \beta^3\\ &
\\ &
C_{2,2,1,1} = 8 N+(6-6 \beta) N^2+(1+4 \beta+\beta^2) N^3+(2 \beta-2 \beta^2) N^4+N^5 \beta^2\\ &
\\ &
C_{2,1,1,1,1} = 12 N^2+(3-3 \beta) N^3+3 N^4 \beta, \ \ \ \ C_{1,1,1,1,1,1} = 15N^3
\end{align*}
\smallskip\\
For odd $k_1 + \ldots + k_m$, correlators vanish due to reflection symmetry of the Gaussian action ${\rm Tr} \ (\phi)^2$. Note, that the above formulas describe correlators with non-vanishing indices $k_i > 0$: every vanishing index $k_i = 0$ simply contributes an overall factor of $N$ to the correlator. This (formulas (\ref{DV1}) and (\ref{NormConst}), (\ref{DV2})), (\ref{DV4})) completes our description of DV expansion for generic potentials $V(z)$. We now turn to discussion of the particular case of 3-Penner potential $V(z) = \alpha_1 \log (z) + \alpha_2 \log( z - q ) + \alpha_3 \log( z - 1 )$.

\section{$\hbar$-expansion for the 3-Penner case}

We are interested in applying the general DV method, described in section 3, to the particular partition function
\begin{align}
Z_{DF} = q^{\dfrac{\alpha_1 \alpha_2}{2 \beta \hbar^2}} \
(1 - q)^{\dfrac{\alpha_2 \alpha_3}{2 \beta \hbar^2}}
\prod\limits_{i = 1}^{N_1} \int\limits_{0}^{q} d z_i \
\prod\limits_{i = N_1+1}^{N_1 + N_2} \int\limits_{0}^{1} d z_i \
\prod\limits_{i < j} (z_j - z_i)^{2 \beta}
\prod\limits_{i} z_i^{\alpha_1/\hbar} (z_i - q)^{\alpha_2/\hbar} (z_i - 1)^{\alpha_3/\hbar}
\label{Dotsenko-Fateev}
\end{align}
i.e, to the particular log-potential 

\begin{align}
V(z) = - \alpha_1 \log (z) - \alpha_2 \log( z - q ) - \alpha_3 \log( z - 1 )
\end{align}
\smallskip\\
This partition function has a DV expansion in powers of $\hbar$, which we denote as $Z^{(DV)}$:
\be
Z^{(DV)} = Z_{U(1)} Z_{0} Z_{\hbar}
\ee
Here the first part is a simple additional prefactor from the definition of Dotsenko-Fateev partition function
\begin{align}
Z_{U(1)} = q^{\dfrac{\alpha_1 \alpha_2}{2 \beta \hbar^2}} \
(1 - q)^{\dfrac{\alpha_2 \alpha_3}{2 \beta \hbar^2}}
\label{QPref}
\end{align}
while the second (the quasiclassical part $Z_0$ ) and the third (the quantum part $Z_{\hbar}$) are given by formulas from section 3 and depend on the critical points -- solutions of the equation of motion

\be
\dfrac{\partial V(z)}{\partial z} = - \dfrac{\alpha_1}{z} - \dfrac{\alpha_2}{ z - q} - \dfrac{\alpha_3}{z - 1} = 0
\ee
\smallskip\\
This equation of motion has just two solutions $z = \mu_1, \mu_2$:

\be
\mu_{1,2} = \dfrac{1}{2} \dfrac{(\alpha_1 + \alpha_2) + (\alpha_1 + \alpha_3) q}{\alpha_1 + \alpha_2 + \alpha_3} \mp \dfrac{1}{2} \dfrac{\sqrt{(\alpha_1 + \alpha_2)^2 - 2(\alpha_1^2 + \alpha_1 \alpha_2+ \alpha_1 \alpha_3 - \alpha_2 \alpha_3) q + (\alpha_1 + \alpha_3)^2 q^2}}{\alpha_1 + \alpha_2 + \alpha_3}
\ee
\smallskip\\
Accordingly, the quasiclassical part takes form
\begin{align}
Z_0 = (\mu_1 - \mu_2)^{2\beta N_1 N_2} \prod\limits_{a = 1}^{2} {\rm Vol}_\beta(N_a) \exp\left( - \dfrac{N_a}{\hbar} V\big(\mu_a) \right) \left( \dfrac{\hbar}{\Delta_a} \right)^{\dfrac{\beta N_a(N_a - 1) + N_a}{2}}
\end{align}
and the quantum part takes form
\begin{align}
Z_{\hbar} = 1 + \sum\limits_{k > 0} H_k \hbar^k
\end{align}
where the quantities $\Delta_1, \Delta_2$ and $H_k$ are given by formulas from section 3 and straightforwardly computable. The quasiclassical part is leading in the regime $\hbar \rightarrow 0$ and non-perturbative in $\hbar$, while the quantum part has a form of power series expansion and is suppressed in the limit $\hbar \rightarrow 0$. However, we are interested not in the DV expansion itself, but rather in comparison with a different expansion in powers of $q$. For this purpose, we need to treat all the quantities of interest (including the critical points $\mu_a$, the second derivatives $\Delta_a$ and higher derivatives) as series in $q$. It is easy to see, that $q$-expansions for the critical points
\be
\mu_1 = \dfrac{\alpha_1}{\alpha_1 + \alpha_2} q - \dfrac{\alpha_1 \alpha_2 \alpha_3}{(\alpha_1 + \alpha_2)^3} q^2 + \ldots
\ee
\be
\mu_2 = \dfrac{\alpha_1 + \alpha_2}{\alpha_1 + \alpha_2 + \alpha_3} + \dfrac{\alpha_2 \alpha_3}{(\alpha_1 + \alpha_2)(\alpha_1 + \alpha_2 + \alpha_3)} q + \dfrac{\alpha_1 \alpha_2 \alpha_3}{(\alpha_1 + \alpha_2)^3} q^2 + \ldots
\ee
imply the following expansions for the three parts of partition function:
\begin{align}
Z_{U(1)} = q^{\deg_{1}} \times \left( 1 + \sum\limits_{k = 1}^{\infty} \sum\limits_{m = -k}^{-1} a_{km} q^k \hbar^{2m} \right)
\end{align}
\begin{align}
Z_0 = C \times \hbar^{\deg_{\hbar}} \times q^{\deg_0} \times \left( 1 + \sum\limits_{k = 1}^{\infty} \sum\limits_{m = -k}^{0} b_{km} q^k \hbar^m \right)
\end{align}
\begin{align}
Z_\hbar = 1 + \sum\limits_{k = 0}^{\infty} \sum\limits_{m = 1}^{\infty} c_{km} q^k \hbar^m
\end{align}
Accordingly, the full partition function posesses the following double $(q, \hbar)$ expansion expansion:

\begin{align}
Z^{(DV)} = C \times \hbar^{\deg_{\hbar}} \times q^{\deg_q} \times \left( \sum\limits_{k = 0}^{\infty} \sum\limits_{m = -2k}^{\infty} Z_{km} q^k \hbar^m \right)
\label{MainExpansion}
\end{align}
\smallskip\\
Complete information about the partition function is encoded in the following data: the overall normalisation constant $C$, the overall degrees $\deg_{\hbar}, \deg_{q}$ and the infinite set of coefficients $\big\{ Z_{km} \big\}$. All these quantities are straightforwardly computable with the quasiclassical DV methods of section 3. The constant $C$ is given by

\begin{align}
\nonumber \log C \ = \ & \sum\limits_{a = 1}^{2} \log {\rm Vol}_\beta(N_a) + \left( \dfrac{1 - \beta}{2} N_1 + \dfrac{\beta}{2} N_1^2 \right) \log ( \alpha_1 \alpha_2 \alpha_3 ) + \dfrac{1}{\hbar} N_1 (\alpha_1 \log \alpha_1 + \alpha_2 \log \alpha_2 + \alpha_3 \log \alpha_3) + \emph{} & \\ \nonumber & \\ \nonumber & \emph{} + \left( 2 \beta N_1 N_2 + \dfrac{3 - 3\beta}{2} N_1 + \dfrac{3 \beta}{2} N_1^2 + \dfrac{1 - \beta}{2} N_2 + \dfrac{\beta}{2} N_2^2 - \dfrac{1}{\hbar}(\alpha_1 + \alpha_2)(N_1 - N_2) \right) \log (\alpha_1 + \alpha_2) - \emph{} & \\ \nonumber & \\ & \emph{} - \left( 2 \beta N_1 N_2 + \dfrac{3 - 3\beta}{2} N_2 + \dfrac{3 \beta}{2} N_2^2 + \dfrac{1}{\hbar}(\alpha_1 + \alpha_2 + \alpha_3) N_2 \right) \log (\alpha_1 + \alpha_2 + \alpha_3)
\end{align}
The overall degrees are given by

\begin{align}
\deg_\hbar = \dfrac{1 - \beta}{2} N_1 + \dfrac{1 - \beta}{2} N_2 + \dfrac{\beta}{2} N_1^2 + \dfrac{\beta}{2} N_2^2
\end{align}
\begin{align}
\deg_q = \dfrac{\alpha_1 \alpha_2}{2\beta\hbar^2} + \beta N_1(N_1 - 1) + N_1 + \dfrac{\alpha_1 + \alpha_2}{\hbar} N_1
\end{align}
\smallskip\\
The first several coefficients $Z_{km}$ are given by

\[
\begin{array}{c|c|c}
Z_{km} & k = 0 & k = 1 \\
\hline
m = -4 & 0 & 0 \\
\hline
m = -3 & 0 & 0  \\
\hline
m = -2 & 0 & a_{11}  \\
\hline
m = -1 & 0 & a_{11} c_{01} + b_{11} \\
\hline
m = 0 & 1 & a_{11} c_{02} + b_{11} c_{01} + b_{10}  \\
\hline
m = 1 & c_{01} & c_{11} + b_{11} c_{02} + a_{11} c_{03} + b_{10} c_{01}  \\
\hline
m = 2 & c_{02} & c_{1 2}+a_{11} c_{0 4}+b_{1 1} c_{0 3}+b_{1 0} c_{0 2}  \\
\hline
\end{array}
\]

\[
\begin{array}{c|c}
Z_{km} & k = 2 \\
\hline
m = -4 & a_{22}  \\
\hline
m = -3 & a_{22} c_{0 1}+a_{11} b_{1 1}  \\
\hline
m = -2 & a_{22} c_{0 2}+a_{11} b_{1 1} c_{0 1}+a_{11} b_{1 0}+b_{2 2}+a_{21}  \\
\hline
m = -1 & a_{1 1} c_{1 1}+a_{1 1} b_{1 1} c_{0 2}+a_{2 2} c_{0 3}+b_{2 1}+c_{0 1} a_{1 1} b_{1 0}+c_{0 1} b_{2 2}+c_{0 1} a_{2 1} \\
\hline
m = 0 & a_{1 1} c_{1 2}+b_{1 1} c_{1 1}+a_{2 2} c_{0 4}+a_{1 1} b_{1 1} c_{0 3}+c_{0 2} a_{1 1} b_{1 0}+c_{0 2} b_{2 2}+c_{0 2} a_{2 1}+b_{2 1} c_{0 1}+b_{2 0} \\
\hline
m = 1 & c_{2 1}+b_{1 1} c_{1 2}+b_{1 0} c_{1 1}+a_{1 1} c_{1 3}+a_{2 2} c_{0 5}+a_{1 1} b_{1 1} c_{0 4}+c_{0 3} a_{1 1} b_{1 0}+c_{0 3} b_{2 2}+c_{0 3} a_{2 1}+b_{2 1} c_{0 2}+b_{2 0} c_{0 1}  \\
\hline
m = 2 & c_{2 2}+b_{1 0} c_{1 2}+a_{1 1} c_{1 4}+b_{1 1} c_{1 3}+b_{2 0} c_{0 2}+b_{2 1} c_{0 3}+c_{0 4} a_{1 1} b_{1 0}+c_{0 4} b_{2 2}+c_{0 4} a_{2 1}+a_{2 2} c_{0 6}+a_{1 1} b_{1 1} c_{0 5} \\
\hline
\end{array}
\]
The first several coefficients $a_{km}$ are given by

\begin{align}
a_{11} = - \dfrac{1}{2} \dfrac{\alpha_2 \alpha_3}{\beta}, \ \ \
a_{21} = - \dfrac{1}{4} \dfrac{\alpha_2 \alpha_3}{\beta}, \ \ \
a_{31} = - \dfrac{1}{6} \dfrac{\alpha_2 \alpha_3}{\beta}, \ \ \
a_{41} = - \dfrac{1}{8} \dfrac{\alpha_2 \alpha_3}{\beta}, \ \ \
a_{51} = - \dfrac{1}{10} \dfrac{\alpha_2 \alpha_3}{\beta}, \ \ \
\end{align}
\begin{align}
a_{22} = \dfrac{1}{8} \dfrac{\alpha_2^2 \alpha_3^2}{\beta^2}, \ \ \
a_{32} = \dfrac{1}{8} \dfrac{\alpha_2^2 \alpha_3^2}{\beta^2}, \ \ \
a_{42} = \dfrac{11}{96} \dfrac{\alpha_2^2 \alpha_3^2}{\beta^2}, \ \ \
a_{52} = \dfrac{5}{48} \dfrac{\alpha_2^2 \alpha_3^2}{\beta^2}
\end{align}
\begin{align}
a_{33} = - \dfrac{1}{48} \dfrac{\alpha_2^3 \alpha_3^3}{\beta^3}, \ \ \
a_{43} = - \dfrac{1}{32} \dfrac{\alpha_2^3 \alpha_3^3}{\beta^3}, \ \ \
a_{53} = - \dfrac{7}{192} \dfrac{\alpha_2^3 \alpha_3^3}{\beta^3}
\end{align}
\begin{align}
a_{44} = \dfrac{1}{384} \dfrac{\alpha_2^4 \alpha_3^4}{\beta^4}, \ \ \
a_{54} = \dfrac{1}{192} \dfrac{\alpha_2^4 \alpha_3^4}{\beta^4}
\end{align}
\begin{align}
a_{55} = - \dfrac{1}{3840} \dfrac{\alpha_2^5 \alpha_3^5}{\beta^5}
\end{align}
The first several coefficients $b_{km}$ are given by
\begin{center}
$b_{10} = \dfrac{1}{(\alpha_{1}+\alpha_{2})^2} \times \Big(\alpha_{2} \alpha_{3} N_{1}+\alpha_{2}^2 N_{2}-\alpha_{1} \alpha_{3} N_{1}+\alpha_{1} \alpha_{2} N_{2}-\alpha_{2} \alpha_{3} N_{1} \beta-\alpha_{2}^2 N_{2} \beta+\alpha_{1} \alpha_{3} N_{1} \beta-\alpha_{1} \alpha_{2} N_{2} \beta-2 \alpha_{2} \alpha_{3} N_{1} N_{2} \beta+\alpha_{2} \alpha_{3} N_{1}^2 \beta+\alpha_{2}^2 N_{2}^2 \beta+2 \alpha_{1} \alpha_{3} N_{1} N_{2} \beta-\alpha_{1} \alpha_{3} N_{1}^2 \beta+\alpha_{1} \alpha_{2} N_{2}^2 \beta+2 \alpha_{1} \alpha_{2} N_{1} N_{2} \beta+2 \alpha_{1}^2 N_{1} N_{2} \beta\Big)$
\end{center}
\begin{center}
$b_{11} = - \dfrac{\alpha_{2} \alpha_{3} N_{2}+\alpha_{2}^2 N_{2}+\alpha_{1} \alpha_{3} N_{1}+\alpha_{1} \alpha_{2} N_{2}}{\alpha_{1}+\alpha_{2}}$
\end{center}
\begin{center}
$b_{20} = \dfrac{1}{2 (\alpha_{1}+\alpha_{2})^4} \times \Big(3 \alpha_{1} \alpha_{2}^2 \alpha_{3} N_{2}^2 \beta+\alpha_{2}^4 N_{2}+3 \alpha_{1}^2 \alpha_{2} \alpha_{3} N_{2}^2 \beta-4 \alpha_{1} \alpha_{2} \alpha_{3}^2 N_{1}^3 \beta^2-12 \alpha_{1} \alpha_{2} \alpha_{3}^2 N_{1} N_{2} \beta+8 \alpha_{1} \alpha_{2} \alpha_{3}^2 N_{1}^2 N_{2} \beta^2-\alpha_{2}^4 N_{2}^2-4 \alpha_{1}^2 \alpha_{3}^2 N_{1}^2 N_{2} \beta^2+5 \alpha_{1}^2 \alpha_{2}^2 N_{2}-2 \alpha_{1}^3 \alpha_{3} N_{1}+2 \alpha_{1}^3 \alpha_{2} N_{2}+4 \alpha_{1} \alpha_{2}^3 N_{2}-\alpha_{1}^2 \alpha_{3}^2 N_{1}-\alpha_{2}^2 \alpha_{3}^2 N_{1}^2-\alpha_{2}^2 \alpha_{3}^2 N_{1}-\alpha_{2}^4 N_{2}^4 \beta^2-\alpha_{2}^4 N_{2}^2 \beta^2-\alpha_{1}^2 \alpha_{3}^2 N_{1}^2-\alpha_{1}^2 \alpha_{2}^2 N_{2}^2-2 \alpha_{1} \alpha_{2}^3 N_{2}^2-2 \alpha_{2}^4 N_{2}^3 \beta+2 \alpha_{2}^4 N_{2}^3 \beta^2-\alpha_{2}^4 N_{2} \beta+\alpha_{1} \alpha_{2} \alpha_{3}^2 N_{1}^2 \beta-\alpha_{1} \alpha_{2} \alpha_{3}^2 N_{2} \beta+\alpha_{1} \alpha_{2} \alpha_{3}^2 N_{2}^2 \beta-5 \alpha_{1} \alpha_{2} \alpha_{3}^2 N_{1} \beta+2 \alpha_{1} \alpha_{2} \alpha_{3}^2 N_{1}^2 \beta^2-2 \alpha_{1}^3 \alpha_{2} N_{2} \beta-4 \alpha_{1} \alpha_{2}^3 N_{2} \beta-2 \alpha_{1} \alpha_{2}^3 N_{2}^2 \beta^2-4 \alpha_{1} \alpha_{2}^3 N_{2}^3 \beta-2 \alpha_{1} \alpha_{2}^3 N_{2}^4 \beta^2+2 \alpha_{1} \alpha_{2} \alpha_{3}^2 N_{1}^2+5 \alpha_{1} \alpha_{2} \alpha_{3}^2 N_{1}+\alpha_{2}^2 \alpha_{3}^2 N_{1}^2 \beta+4 \alpha_{1} \alpha_{2}^3 N_{2}^3 \beta^2+7 \alpha_{1}^2 \alpha_{2}^2 N_{2}^2 \beta+2 \alpha_{1}^3 \alpha_{2} N_{2}^2 \beta+8 \alpha_{1} \alpha_{2}^3 N_{2}^2 \beta+2 \alpha_{1}^2 \alpha_{2}^2 N_{2}^3 \beta^2+3 \alpha_{1}^2 \alpha_{2} \alpha_{3} N_{2}+\alpha_{2}^2 \alpha_{3}^2 N_{1} \beta+\alpha_{1} \alpha_{2} \alpha_{3}^2 N_{2}-\alpha_{2}^2 \alpha_{3}^2 N_{1}^2 \beta^2+\alpha_{1}^2 \alpha_{3}^2 N_{1} \beta+2 \alpha_{1}^4 N_{1} N_{2} \beta-2 \alpha_{1}^2 \alpha_{3}^2 N_{1}^3 \beta-\alpha_{1}^2 \alpha_{3}^2 N_{1}^4 \beta^2+2 \alpha_{1}^2 \alpha_{2} \alpha_{3} N_{1} N_{2} \beta^2+4 \alpha_{1} \alpha_{2} \alpha_{3}^2 N_{1}^3 \beta-8 \alpha_{1} \alpha_{2} \alpha_{3}^2 N_{1}^3 N_{2} \beta^2+4 \alpha_{2}^3 \alpha_{3} N_{1} N_{2}^3 \beta^2+8 \alpha_{1} \alpha_{2} \alpha_{3}^2 N_{1}^2 N_{2}^2 \beta^2-4 \alpha_{1} \alpha_{2}^3 N_{1} N_{2}^2 \beta-4 \alpha_{1} \alpha_{2}^2 \alpha_{3} N_{1}^3 N_{2} \beta^2+2 \alpha_{1}^2 \alpha_{2} \alpha_{3} N_{1}^2 N_{2}^2 \beta^2+2 \alpha_{2}^3 \alpha_{3} N_{1} N_{2}^2 \beta-4 \alpha_{1}^2 \alpha_{2} \alpha_{3} N_{1} N_{2}^3 \beta^2-2 \alpha_{1}^2 \alpha_{2} \alpha_{3} N_{1} N_{2}^2 \beta+\alpha_{1}^2 \alpha_{2} \alpha_{3} N_{1}^2 \beta-2 \alpha_{1}^2 \alpha_{2} \alpha_{3} N_{1}^2 N_{2} \beta^2+2 \alpha_{2}^3 \alpha_{3} N_{1}^2 N_{2} \beta^2+2 \alpha_{1}^2 \alpha_{2}^2 N_{1} N_{2} \beta-2 \alpha_{2}^3 \alpha_{3} N_{1} N_{2}^2 \beta^2-4 \alpha_{2}^2 \alpha_{3}^2 N_{1}^2 N_{2} \beta^2-4 \alpha_{1}^3 \alpha_{3} N_{1}^2 N_{2} \beta^2+4 \alpha_{2}^3 \alpha_{3} N_{1} N_{2} \beta+2 \alpha_{2}^2 \alpha_{3}^2 N_{1} N_{2} \beta+4 \alpha_{1}^3 \alpha_{3} N_{1} N_{2} \beta+4 \alpha_{1}^3 \alpha_{2} N_{1} N_{2} \beta-2 \alpha_{2}^3 \alpha_{3} N_{1} N_{2} \beta^2+4 \alpha_{1}^3 \alpha_{2} N_{1} N_{2}^2 \beta^2-4 \alpha_{1} \alpha_{2}^2 \alpha_{3} N_{1}^2 N_{2} \beta-4 \alpha_{1}^3 \alpha_{2} N_{1} N_{2}^3 \beta^2-8 \alpha_{1}^3 \alpha_{2} N_{1}^2 N_{2}^2 \beta^2+4 \alpha_{1}^2 \alpha_{3}^2 N_{1}^3 N_{2} \beta^2+2 \alpha_{1}^2 \alpha_{2} \alpha_{3} N_{1} N_{2}-4 \alpha_{1}^2 \alpha_{3}^2 N_{1}^2 N_{2}^2 \beta^2-4 \alpha_{1}^3 \alpha_{2} N_{1} N_{2}^2 \beta-4 \alpha_{2}^2 \alpha_{3}^2 N_{1}^2 N_{2}^2 \beta^2+2 \alpha_{1}^2 \alpha_{3}^2 N_{1} N_{2} \beta+4 \alpha_{1} \alpha_{2}^2 \alpha_{3} N_{1}^2 N_{2} \beta^2+4 \alpha_{1} \alpha_{2}^3 N_{1} N_{2}^2 \beta^2+8 \alpha_{1}^2 \alpha_{2}^2 N_{1} N_{2}^2 \beta^2-12 \alpha_{1}^2 \alpha_{2} \alpha_{3} N_{1} N_{2} \beta-12 \alpha_{1} \alpha_{2}^2 \alpha_{3} N_{1} N_{2} \beta-3 \alpha_{1} \alpha_{2}^2 \alpha_{3} N_{2} \beta+3 \alpha_{1} \alpha_{2}^2 \alpha_{3} N_{1}^2 \beta-3 \alpha_{1} \alpha_{2}^2 \alpha_{3} N_{1} \beta-\alpha_{1}^2 \alpha_{2} \alpha_{3} N_{1} \beta+2 \alpha_{1} \alpha_{2} \alpha_{3}^2 N_{1}^4 \beta^2+8 \alpha_{1} \alpha_{2}^2 \alpha_{3} N_{1}^2 N_{2}^2 \beta^2-4 \alpha_{1} \alpha_{2}^3 N_{1} N_{2}^3 \beta^2-8 \alpha_{1} \alpha_{2} \alpha_{3}^2 N_{1}^2 N_{2} \beta-2 \alpha_{2}^3 \alpha_{3} N_{1}^2 N_{2}^2 \beta^2-4 \alpha_{1}^2 \alpha_{2}^2 N_{1}^2 N_{2}^2 \beta^2-8 \alpha_{1}^2 \alpha_{2}^2 N_{1} N_{2}^3 \beta^2-\alpha_{1}^2 \alpha_{2}^2 N_{2}^4 \beta^2-2 \alpha_{2}^2 \alpha_{3}^2 N_{1}^3 \beta-4 \alpha_{1}^4 N_{1}^2 N_{2}^2 \beta^2-2 \alpha_{1}^2 \alpha_{2}^2 N_{2}^3 \beta-\alpha_{2}^2 \alpha_{3}^2 N_{1}^4 \beta^2-2 \alpha_{2}^3 \alpha_{3} N_{1} N_{2}+4 \alpha_{1}^3 \alpha_{3} N_{1}^3 N_{2} \beta^2+2 \alpha_{1}^2 \alpha_{2} \alpha_{3} N_{1}^2 N_{2} \beta+4 \alpha_{1}^3 \alpha_{3} N_{1}^2 N_{2} \beta-8 \alpha_{1}^3 \alpha_{3} N_{1}^2 N_{2}^2 \beta^2-2 \alpha_{2}^3 \alpha_{3} N_{1}^2 N_{2} \beta+4 \alpha_{2}^2 \alpha_{3}^2 N_{1}^3 N_{2} \beta^2+4 \alpha_{2}^2 \alpha_{3}^2 N_{1}^2 N_{2} \beta+4 \alpha_{1}^2 \alpha_{3}^2 N_{1}^2 N_{2} \beta-8 \alpha_{1}^2 \alpha_{2}^2 N_{1} N_{2}^2 \beta+2 \alpha_{1}^2 \alpha_{2} \alpha_{3} N_{1} N_{2}^2 \beta^2-3 \alpha_{1}^2 \alpha_{2} \alpha_{3} N_{2} \beta+3 \alpha_{2}^4 N_{2}^2 \beta+2 \alpha_{1}^2 \alpha_{3}^2 N_{1}^3 \beta^2+\alpha_{1}^2 \alpha_{3}^2 N_{1}^2 \beta-\alpha_{1}^2 \alpha_{3}^2 N_{1}^2 \beta^2+3 \alpha_{1} \alpha_{2}^2 \alpha_{3} N_{2}+\alpha_{1}^2 \alpha_{2} \alpha_{3} N_{1}+2 \alpha_{1}^3 \alpha_{3} N_{1} \beta-2 \alpha_{1}^3 \alpha_{3} N_{1}^2 \beta+3 \alpha_{1} \alpha_{2}^2 \alpha_{3} N_{1}+2 \alpha_{2}^2 \alpha_{3}^2 N_{1}^3 \beta^2-\alpha_{1}^2 \alpha_{2}^2 N_{2}^2 \beta^2-5 \alpha_{1}^2 \alpha_{2}^2 N_{2} \beta\Big)$
\end{center}
\begin{center}
$b_{21} = \dfrac{1}{2 (\alpha_{1}+\alpha_{2})^3} \times \Big(2 \alpha_{1} \alpha_{2}^2 \alpha_{3} N_{2}^2 \beta+\alpha_{2}^4 N_{2}-2 \alpha_{1} \alpha_{2} \alpha_{3}^2 N_{1} N_{2} \beta-2 \alpha_{2}^4 N_{2}^2+3 \alpha_{1}^2 \alpha_{2}^2 N_{2}+\alpha_{1}^3 \alpha_{3} N_{1}+\alpha_{1}^3 \alpha_{2} N_{2}+3 \alpha_{1} \alpha_{2}^3 N_{2}+2 \alpha_{1}^2 \alpha_{3}^2 N_{1}^2-2 \alpha_{1}^2 \alpha_{2}^2 N_{2}^2-4 \alpha_{1} \alpha_{2}^3 N_{2}^2-2 \alpha_{2}^4 N_{2}^3 \beta+\alpha_{2}^3 \alpha_{3} N_{2}-2 \alpha_{2}^3 \alpha_{3} N_{2}^2-2 \alpha_{1} \alpha_{2}^2 \alpha_{3} N_{1} N_{2}+4 \alpha_{2}^2 \alpha_{3}^2 N_{1} N_{2}^2 \beta-4 \alpha_{1} \alpha_{2} \alpha_{3}^2 N_{1} N_{2}^2 \beta+2 \alpha_{1} \alpha_{2} \alpha_{3}^2 N_{1} N_{2}-6 \alpha_{1} \alpha_{2}^2 \alpha_{3} N_{1} N_{2}^2 \beta-2 \alpha_{1} \alpha_{2}^2 \alpha_{3} N_{2}^3 \beta+2 \alpha_{1} \alpha_{2} \alpha_{3}^2 N_{1}^2 \beta-4 \alpha_{1} \alpha_{2}^3 N_{2}^3 \beta-2 \alpha_{1} \alpha_{2} \alpha_{3}^2 N_{1}^2-\alpha_{1} \alpha_{2} \alpha_{3}^2 N_{1}+2 \alpha_{1}^2 \alpha_{2}^2 N_{2}^2 \beta+4 \alpha_{1} \alpha_{2}^3 N_{2}^2 \beta+2 \alpha_{1}^2 \alpha_{2} \alpha_{3} N_{2}+\alpha_{1} \alpha_{2} \alpha_{3}^2 N_{2}+2 \alpha_{1}^2 \alpha_{3}^2 N_{1}^3 \beta-2 \alpha_{1} \alpha_{2} \alpha_{3}^2 N_{1}^3 \beta-4 \alpha_{1} \alpha_{2}^3 N_{1} N_{2}^2 \beta+4 \alpha_{2}^3 \alpha_{3} N_{1} N_{2}^2 \beta-10 \alpha_{1}^2 \alpha_{2} \alpha_{3} N_{1} N_{2}^2 \beta+2 \alpha_{2}^3 \alpha_{3} N_{1} N_{2} \beta+2 \alpha_{2}^2 \alpha_{3}^2 N_{1} N_{2} \beta-4 \alpha_{1}^3 \alpha_{2} N_{1} N_{2}^2 \beta+2 \alpha_{1} \alpha_{2}^2 \alpha_{3} N_{1} N_{2} \beta+6 \alpha_{1} \alpha_{2} \alpha_{3}^2 N_{1}^2 N_{2} \beta-2 \alpha_{1}^2 \alpha_{2}^2 N_{2}^3 \beta-2 \alpha_{2}^3 \alpha_{3} N_{1} N_{2}-2 \alpha_{1}^2 \alpha_{2} \alpha_{3} N_{1}^2 N_{2} \beta-4 \alpha_{1}^3 \alpha_{3} N_{1}^2 N_{2} \beta-2 \alpha_{2}^3 \alpha_{3} N_{1}^2 N_{2} \beta-2 \alpha_{2}^2 \alpha_{3}^2 N_{1}^2 N_{2} \beta-4 \alpha_{1}^2 \alpha_{3}^2 N_{1}^2 N_{2} \beta-8 \alpha_{1}^2 \alpha_{2}^2 N_{1} N_{2}^2 \beta+2 \alpha_{2}^4 N_{2}^2 \beta-2 \alpha_{1} \alpha_{2}^2 \alpha_{3} N_{2}^2-2 \alpha_{2}^2 \alpha_{3}^2 N_{1} N_{2}+2 \alpha_{2}^3 \alpha_{3} N_{2}^2 \beta-2 \alpha_{2}^3 \alpha_{3} N_{2}^3 \beta-2 \alpha_{1}^2 \alpha_{3}^2 N_{1}^2 \beta+3 \alpha_{1} \alpha_{2}^2 \alpha_{3} N_{2}+\alpha_{1}^2 \alpha_{2} \alpha_{3} N_{1}\Big)$
\end{center}
\begin{center}
$b_{22} = - \dfrac{(\alpha_{2} \alpha_{3} N_{2}+\alpha_{2}^2 N_{2}+\alpha_{1} \alpha_{3} N_{1}+\alpha_{1} \alpha_{2} N_{2})^2}{2 (\alpha_{1}+\alpha_{2})^4}$
\end{center}
The first several coefficients $c_{km}$ are given by
\begin{center}
$c_{01} = \dfrac{1}{12 \alpha_{3} \alpha_{2} \alpha_{1} (\alpha_{1}+\alpha_{2}) (\alpha_{1}+\alpha_{2}+\alpha_{3})} \times \Big( \alpha_{1}^2 \alpha_{3}^2 N_{1}-3 \alpha_{1}^2 \alpha_{3}^2 N_{1}^2 \beta^2+2 \alpha_{1}^2 \alpha_{3}^2 N_{1}^3 \beta^2+\alpha_{2}^3 \alpha_{3} N_{1}-21 \alpha_{1} \alpha_{2}^2 \alpha_{3} N_{2}^2 \beta-3 \alpha_{1} \alpha_{2} \alpha_{3}^2 N_{2} \beta+3 \alpha_{1} \alpha_{2} \alpha_{3}^2 N_{2}^2 \beta+2 \alpha_{1} \alpha_{2} \alpha_{3}^2 N_{2}^3 \beta^2-3 \alpha_{1} \alpha_{2} \alpha_{3}^2 N_{2}^2 \beta^2-11 \alpha_{1} \alpha_{2} \alpha_{3}^2 N_{1} \beta^2+21 \alpha_{1} \alpha_{2} \alpha_{3}^2 N_{1} \beta+21 \alpha_{1} \alpha_{2} \alpha_{3}^2 N_{1}^2 \beta^2+21 \alpha_{1} \alpha_{2}^2 \alpha_{3} N_{2}^2 \beta^2+2 \alpha_{2}^2 \alpha_{3}^2 N_{1}^3 \beta^2-6 \alpha_{1}^2 \alpha_{2}^2 N_{2}^2 \beta^2+2 \alpha_{1}^2 \alpha_{2}^2 N_{2} \beta^2+2 \alpha_{2}^3 \alpha_{3} N_{1}^3 \beta^2+3 \alpha_{1}^3 \alpha_{3} N_{1}^2 \beta-10 \alpha_{1} \alpha_{2}^2 \alpha_{3} N_{1}-3 \alpha_{2}^3 \alpha_{3} N_{1} \beta-3 \alpha_{1}^3 \alpha_{3} N_{1}^2 \beta^2-3 \alpha_{1}^3 \alpha_{3} N_{1} \beta+2 \alpha_{1}^3 \alpha_{3} N_{1}^3 \beta^2-3 \alpha_{1} \alpha_{2}^3 N_{2}^2 \beta^2+3 \alpha_{2}^2 \alpha_{3}^2 N_{1}^2 \beta+2 \alpha_{1} \alpha_{2}^3 N_{2}^3 \beta^2-6 \alpha_{1}^2 \alpha_{2}^2 N_{2} \beta-3 \alpha_{1} \alpha_{2}^3 N_{2} \beta-3 \alpha_{1}^3 \alpha_{2} N_{2}^2 \beta^2-3 \alpha_{1}^3 \alpha_{2} N_{2} \beta+4 \alpha_{1}^2 \alpha_{2}^2 N_{2}^3 \beta^2+3 \alpha_{2}^3 \alpha_{3} N_{1}^2 \beta+2 \alpha_{1}^3 \alpha_{2} N_{2}^3 \beta^2-18 \alpha_{1}^2 \alpha_{2} \alpha_{3} N_{1}^2 \beta+3 \alpha_{1}^3 \alpha_{2} N_{2}^2 \beta+\alpha_{2}^2 \alpha_{3}^2 N_{1}+24 \alpha_{1}^2 \alpha_{2} \alpha_{3} N_{1} N_{2} \beta^2-11 \alpha_{1}^2 \alpha_{2} \alpha_{3} N_{2} \beta^2-3 \alpha_{2}^2 \alpha_{3}^2 N_{1}^2 \beta^2-3 \alpha_{1}^2 \alpha_{3}^2 N_{1} \beta-3 \alpha_{2}^3 \alpha_{3} N_{1}^2 \beta^2+\alpha_{2}^2 \alpha_{3}^2 N_{1} \beta^2+\alpha_{1} \alpha_{2}^3 N_{2} \beta^2+\alpha_{1}^3 \alpha_{2} N_{2} \beta^2+\alpha_{2}^3 \alpha_{3} N_{1} \beta^2-3 \alpha_{2}^2 \alpha_{3}^2 N_{1} \beta+\alpha_{1}^3 \alpha_{3} N_{1} \beta^2+\alpha_{1}^2 \alpha_{3}^2 N_{1} \beta^2-11 \alpha_{1} \alpha_{2}^2 \alpha_{3} N_{2} \beta^2-8 \alpha_{1} \alpha_{2}^2 \alpha_{3} N_{1}^3 \beta^2+21 \alpha_{1} \alpha_{2}^2 \alpha_{3} N_{2} \beta+24 \alpha_{1} \alpha_{2}^2 \alpha_{3} N_{1} N_{2} \beta^2-10 \alpha_{1} \alpha_{2}^2 \alpha_{3} N_{1} \beta^2-24 \alpha_{1}^2 \alpha_{2} \alpha_{3} N_{1} N_{2}^2 \beta^2-24 \alpha_{1}^2 \alpha_{2} \alpha_{3} N_{1} N_{2} \beta-10 \alpha_{1} \alpha_{2}^2 \alpha_{3} N_{2}^3 \beta^2+21 \alpha_{1}^2 \alpha_{2} \alpha_{3} N_{2}^2 \beta^2+\alpha_{1} \alpha_{2} \alpha_{3}^2 N_{2} \beta^2-8 \alpha_{1}^2 \alpha_{2} \alpha_{3} N_{1}^3 \beta^2+18 \alpha_{1}^2 \alpha_{2} \alpha_{3} N_{1}^2 \beta^2+18 \alpha_{1}^2 \alpha_{2} \alpha_{3} N_{1} \beta-10 \alpha_{1}^2 \alpha_{2} \alpha_{3} N_{1} \beta^2-18 \alpha_{1} \alpha_{2}^2 \alpha_{3} N_{1}^2 \beta+18 \alpha_{1} \alpha_{2}^2 \alpha_{3} N_{1} \beta-10 \alpha_{1} \alpha_{2} \alpha_{3}^2 N_{1}^3 \beta^2+24 \alpha_{1} \alpha_{2} \alpha_{3}^2 N_{1}^2 N_{2} \beta^2+12 \alpha_{1} \alpha_{2} \alpha_{3}^2 N_{1} N_{2} \beta-21 \alpha_{1}^2 \alpha_{2} \alpha_{3} N_{2}^2 \beta-10 \alpha_{1}^2 \alpha_{2} \alpha_{3} N_{2}^3 \beta^2+18 \alpha_{1} \alpha_{2}^2 \alpha_{3} N_{1}^2 \beta^2-24 \alpha_{1} \alpha_{2}^2 \alpha_{3} N_{1} N_{2}^2 \beta^2+21 \alpha_{1}^2 \alpha_{2} \alpha_{3} N_{2} \beta-12 \alpha_{1} \alpha_{2} \alpha_{3}^2 N_{1} N_{2} \beta^2+12 \alpha_{1} \alpha_{2} \alpha_{3}^2 N_{1} N_{2}^2 \beta^2+2 \alpha_{1}^2 \alpha_{2}^2 N_{2}+\alpha_{1}^3 \alpha_{3} N_{1}+\alpha_{1}^3 \alpha_{2} N_{2}+\alpha_{1} \alpha_{2}^3 N_{2}-21 \alpha_{1} \alpha_{2} \alpha_{3}^2 N_{1}^2 \beta-11 \alpha_{1} \alpha_{2} \alpha_{3}^2 N_{1}+6 \alpha_{1}^2 \alpha_{2}^2 N_{2}^2 \beta+3 \alpha_{1} \alpha_{2}^3 N_{2}^2 \beta-11 \alpha_{1}^2 \alpha_{2} \alpha_{3} N_{2}+\alpha_{1} \alpha_{2} \alpha_{3}^2 N_{2}-24 \alpha_{1} \alpha_{2}^2 \alpha_{3} N_{1} N_{2} \beta+3 \alpha_{1}^2 \alpha_{3}^2 N_{1}^2 \beta-11 \alpha_{1} \alpha_{2}^2 \alpha_{3} N_{2}-10 \alpha_{1}^2 \alpha_{2} \alpha_{3} N_{1} \Big)$
\end{center}
\begin{center}
$c_{11} = - \dfrac{2 N_{1} (\alpha_{1} - \alpha_{2})}{(\alpha_{1}+\alpha_{2})^3} \times \Big( \alpha_{3}-2 \alpha_{3} \beta+\alpha_{3} \beta^2-\alpha_{3} N_{2} \beta+2 \alpha_{3} N_{1} \beta+\alpha_{3} N_{2} \beta^2-2 \alpha_{3} N_{1} \beta^2-3 \alpha_{3} N_{1} N_{2} \beta^2+\alpha_{3} N_{1}^2 \beta^2+\alpha_{2} N_{2}^2 \beta^2-\alpha_{2} N_{1} N_{2} \beta^2+\alpha_{1} N_{2}^2 \beta^2-\alpha_{1} N_{1} N_{2} \beta^2 \Big)$
\end{center}
\begin{center}
$c_{21} = \dfrac{1}{2(\alpha_{1}+\alpha_{2})^5} \times \Big( 12 \alpha_{1} \alpha_{2}^3 N_{1} N_{2}^2 \beta^2+20 \alpha_{1}^2 \alpha_{2}^2 N_{1} N_{2}^2 \beta^2-6 \alpha_{2}^2 \alpha_{3}^2 N_{1} N_{2} \beta^2-6 \alpha_{1}^2 \alpha_{3}^2 N_{1}+12 \alpha_{1}^2 \alpha_{3}^2 N_{1}^2 \beta^2-6 \alpha_{1}^2 \alpha_{3}^2 N_{1}^3 \beta^2-2 \alpha_{2}^3 \alpha_{3} N_{1}-5 \alpha_{1} \alpha_{2}^2 \alpha_{3} N_{2}^2 \beta+\alpha_{1} \alpha_{2} \alpha_{3}^2 N_{2} \beta-\alpha_{1} \alpha_{2} \alpha_{3}^2 N_{2}^2 \beta+\alpha_{1} \alpha_{2} \alpha_{3}^2 N_{2}^2 \beta^2+21 \alpha_{1} \alpha_{2} \alpha_{3}^2 N_{1} \beta^2-41 \alpha_{1} \alpha_{2} \alpha_{3}^2 N_{1} \beta-41 \alpha_{1} \alpha_{2} \alpha_{3}^2 N_{1}^2 \beta^2+5 \alpha_{1} \alpha_{2}^2 \alpha_{3} N_{2}^2 \beta^2-6 \alpha_{2}^2 \alpha_{3}^2 N_{1}^3 \beta^2+8 \alpha_{1}^2 \alpha_{2}^2 N_{2}^2 \beta^2-4 \alpha_{1}^2 \alpha_{2}^2 N_{2} \beta^2-2 \alpha_{2}^3 \alpha_{3} N_{1}^3 \beta^2-12 \alpha_{1}^3 \alpha_{3} N_{1}^2 \beta+9 \alpha_{1} \alpha_{2}^2 \alpha_{3} N_{1}+4 \alpha_{2}^3 \alpha_{3} N_{1} \beta+12 \alpha_{1}^3 \alpha_{3} N_{1}^2 \beta^2+12 \alpha_{1}^3 \alpha_{3} N_{1} \beta-6 \alpha_{1}^3 \alpha_{3} N_{1}^3 \beta^2+4 \alpha_{1} \alpha_{2}^3 N_{2}^2 \beta^2-12 \alpha_{2}^2 \alpha_{3}^2 N_{1}^2 \beta-2 \alpha_{1} \alpha_{2}^3 N_{2}^3 \beta^2+8 \alpha_{1}^2 \alpha_{2}^2 N_{2} \beta+4 \alpha_{1} \alpha_{2}^3 N_{2} \beta+4 \alpha_{1}^3 \alpha_{2} N_{2}^2 \beta^2+4 \alpha_{1}^3 \alpha_{2} N_{2} \beta-4 \alpha_{1}^2 \alpha_{2}^2 N_{2}^3 \beta^2-4 \alpha_{2}^3 \alpha_{3} N_{1}^2 \beta-2 \alpha_{1}^3 \alpha_{2} N_{2}^3 \beta^2+9 \alpha_{1}^2 \alpha_{2} \alpha_{3} N_{1}^2 \beta-4 \alpha_{1}^3 \alpha_{2} N_{2}^2 \beta-6 \alpha_{2}^2 \alpha_{3}^2 N_{1}+2 \alpha_{1}^2 \alpha_{2} \alpha_{3} N_{1} N_{2} \beta^2-3 \alpha_{1}^2 \alpha_{2} \alpha_{3} N_{2} \beta^2+12 \alpha_{2}^2 \alpha_{3}^2 N_{1}^2 \beta^2+12 \alpha_{1}^2 \alpha_{3}^2 N_{1} \beta+4 \alpha_{2}^3 \alpha_{3} N_{1}^2 \beta^2-6 \alpha_{2}^2 \alpha_{3}^2 N_{1} \beta^2-2 \alpha_{1} \alpha_{2}^3 N_{2} \beta^2-2 \alpha_{1}^3 \alpha_{2} N_{2} \beta^2-2 \alpha_{2}^3 \alpha_{3} N_{1} \beta^2+12 \alpha_{2}^2 \alpha_{3}^2 N_{1} \beta-6 \alpha_{1}^3 \alpha_{3} N_{1} \beta^2-6 \alpha_{1}^2 \alpha_{3}^2 N_{1} \beta^2-3 \alpha_{1} \alpha_{2}^2 \alpha_{3} N_{2} \beta^2+8 \alpha_{1} \alpha_{2}^2 \alpha_{3} N_{1}^3 \beta^2+5 \alpha_{1} \alpha_{2}^2 \alpha_{3} N_{2} \beta+6 \alpha_{1} \alpha_{2}^2 \alpha_{3} N_{1} N_{2} \beta^2+9 \alpha_{1} \alpha_{2}^2 \alpha_{3} N_{1} \beta^2+30 \alpha_{1}^2 \alpha_{2} \alpha_{3} N_{1} N_{2}^2 \beta^2-2 \alpha_{1}^2 \alpha_{2} \alpha_{3} N_{1} N_{2} \beta-2 \alpha_{1} \alpha_{2}^2 \alpha_{3} N_{2}^3 \beta^2+5 \alpha_{1}^2 \alpha_{2} \alpha_{3} N_{2}^2 \beta^2-\alpha_{1} \alpha_{2} \alpha_{3}^2 N_{2} \beta^2+4 \alpha_{1}^2 \alpha_{2} \alpha_{3} N_{1}^3 \beta^2-9 \alpha_{1}^2 \alpha_{2} \alpha_{3} N_{1}^2 \beta^2-9 \alpha_{1}^2 \alpha_{2} \alpha_{3} N_{1} \beta+5 \alpha_{1}^2 \alpha_{2} \alpha_{3} N_{1} \beta^2+17 \alpha_{1} \alpha_{2}^2 \alpha_{3} N_{1}^2 \beta-17 \alpha_{1} \alpha_{2}^2 \alpha_{3} N_{1} \beta+20 \alpha_{1} \alpha_{2} \alpha_{3}^2 N_{1}^3 \beta^2-74 \alpha_{1} \alpha_{2} \alpha_{3}^2 N_{1}^2 N_{2} \beta^2-20 \alpha_{1} \alpha_{2} \alpha_{3}^2 N_{1} N_{2} \beta-5 \alpha_{1}^2 \alpha_{2} \alpha_{3} N_{2}^2 \beta-2 \alpha_{1}^2 \alpha_{2} \alpha_{3} N_{2}^3 \beta^2-17 \alpha_{1} \alpha_{2}^2 \alpha_{3} N_{1}^2 \beta^2+30 \alpha_{1} \alpha_{2}^2 \alpha_{3} N_{1} N_{2}^2 \beta^2+5 \alpha_{1}^2 \alpha_{2} \alpha_{3} N_{2} \beta+20 \alpha_{1} \alpha_{2} \alpha_{3}^2 N_{1} N_{2} \beta^2+14 \alpha_{1} \alpha_{2} \alpha_{3}^2 N_{1} N_{2}^2 \beta^2-4 \alpha_{1}^2 \alpha_{2}^2 N_{2}-6 \alpha_{1}^3 \alpha_{3} N_{1}-2 \alpha_{1}^3 \alpha_{2} N_{2}-2 \alpha_{1} \alpha_{2}^3 N_{2}+41 \alpha_{1} \alpha_{2} \alpha_{3}^2 N_{1}^2 \beta+21 \alpha_{1} \alpha_{2} \alpha_{3}^2 N_{1}-8 \alpha_{1}^2 \alpha_{2}^2 N_{2}^2 \beta-4 \alpha_{1} \alpha_{2}^3 N_{2}^2 \beta-3 \alpha_{1}^2 \alpha_{2} \alpha_{3} N_{2}-\alpha_{1} \alpha_{2} \alpha_{3}^2 N_{2}+2 \alpha_{2}^3 \alpha_{3} N_{1} N_{2} \beta+6 \alpha_{2}^2 \alpha_{3}^2 N_{1} N_{2} \beta-6 \alpha_{1} \alpha_{2}^2 \alpha_{3} N_{1} N_{2} \beta-4 \alpha_{1}^4 N_{1} N_{2}^2 \beta^2+4 \alpha_{1}^4 N_{1}^2 N_{2} \beta^2+20 \alpha_{1}^2 \alpha_{3}^2 N_{1}^2 N_{2} \beta^2+2 \alpha_{1} \alpha_{2}^3 N_{1} N_{2} \beta-4 \alpha_{1}^2 \alpha_{2}^2 N_{1} N_{2} \beta^2-2 \alpha_{2}^2 \alpha_{3}^2 N_{1} N_{2}^2 \beta^2-6 \alpha_{2}^3 \alpha_{3} N_{1} N_{2}^2 \beta^2+20 \alpha_{2}^2 \alpha_{3}^2 N_{1}^2 N_{2} \beta^2+12 \alpha_{2}^3 \alpha_{3} N_{1}^2 N_{2} \beta^2-6 \alpha_{1}^2 \alpha_{3}^2 N_{1} N_{2} \beta^2+4 \alpha_{1}^2 \alpha_{2}^2 N_{1} N_{2} \beta-36 \alpha_{1}^2 \alpha_{2} \alpha_{3} N_{1}^2 N_{2} \beta^2-8 \alpha_{1}^2 \alpha_{2}^2 N_{1}^2 N_{2} \beta^2-48 \alpha_{1} \alpha_{2}^2 \alpha_{3} N_{1}^2 N_{2} \beta^2+24 \alpha_{1}^3 \alpha_{3} N_{1}^2 N_{2} \beta^2+6 \alpha_{1}^3 \alpha_{3} N_{1} N_{2} \beta-6 \alpha_{1}^3 \alpha_{3} N_{1} N_{2} \beta^2+2 \alpha_{1}^3 \alpha_{2} N_{1} N_{2} \beta-2 \alpha_{2}^3 \alpha_{3} N_{1} N_{2} \beta^2-6 \alpha_{1}^3 \alpha_{3} N_{1} N_{2}^2 \beta^2-2 \alpha_{1}^3 \alpha_{2} N_{1} N_{2} \beta^2+2 \alpha_{1}^3 \alpha_{2} N_{1}^2 N_{2} \beta^2+4 \alpha_{1}^3 \alpha_{2} N_{1} N_{2}^2 \beta^2+6 \alpha_{1}^2 \alpha_{3}^2 N_{1} N_{2} \beta-12 \alpha_{1}^2 \alpha_{3}^2 N_{1}^2 \beta-3 \alpha_{1} \alpha_{2}^2 \alpha_{3} N_{2}+5 \alpha_{1}^2 \alpha_{2} \alpha_{3} N_{1}-2 \alpha_{1} \alpha_{2}^3 N_{1} N_{2} \beta^2-6 \alpha_{1} \alpha_{2}^3 N_{1}^2 N_{2} \beta^2-2 \alpha_{1}^2 \alpha_{3}^2 N_{1} N_{2}^2 \beta^2 \Big)$
\end{center}
These somewhat lengthy expressions describe the first several orders of double $(\hbar,q)$ expansion of $Z_{DV}$. We now proceed to derive this expansion from a different point of view: namely, starting from the contour-integration method and the corresponding conformal-field-theory $q$-expansion.

\section{$q$-expansion for the 3-Penner case. Comparison}

On the other side, the partition function (\ref{Dotsenko-Fateev}) is known \cite{MaMoAGT2,MaMoAGT3} to posess an expansion in powers of $q$
\begin{align}
Z^{(CFT)} = q^{\deg} \times C_{N_1}\left(\dfrac{\alpha_1}{\hbar}, \dfrac{\alpha_2}{\hbar}\right) \ C_{N_2}\left(\dfrac{\alpha_1 + \alpha_2}{\hbar} + 2 \beta N_1, \dfrac{\alpha_3}{\hbar}\right) \times \left( 1 + \sum\limits_{k = 1}^{\infty} B_k q^k \right)
\end{align}
with the overall degree
\be
\deg = \dfrac{\alpha_1 \alpha_2}{2\beta\hbar^2} + \beta N_1(N_1 - 1) + N_1 + \dfrac{\alpha_1 + \alpha_2}{\hbar} N_1
\ee
the normalisation constants
\be
C_{N}(x,y) = \prod\limits_{k = 1}^{N} \dfrac{\Gamma(x + 1 + \beta(k-1))
\Gamma(y + 1 + \beta(k-1))\Gamma(1 + \beta k)}
{\Gamma(x + y + 2 + (N+k-2)\beta)\Gamma(\beta + 1)}
\ee
and the first coefficient
\begin{align}
\nonumber B_1 \ = \ & -(2 \beta N_1-2 N_1 \beta^2+2 \beta^2 N_1^2+2 N_1 \beta \alpha_{1}+2 \alpha_{2}-2 \beta \alpha_{2}+2 \beta N_1 \alpha_{2}+\alpha_{2} \alpha_{1}+\alpha_{2}^2) \times \emph{} & \\ \nonumber & \\ \nonumber & \emph{}
\times (4 N_2 N_1 \beta^2+2 \beta N_1 \alpha_{3}+2 \beta^2 N_2^2-2 \beta^2 N_2+2 \beta N_2 \alpha_{2}+2 \beta N_2 \alpha_{1}+2 N_2 \alpha_{3} \beta+2 \beta N_2+\alpha_{3} \alpha_{1}+\alpha_{3} \alpha_{2}) \times \emph{} & \\ \nonumber & \\ & \emph{} \times (2\beta)^{-1} (2 \beta N_1+\alpha_{1}+\alpha_{2})^{-1} (2 \beta N_1+\alpha_{1}+\alpha_{2}-2 \beta+2)^{-1}
\end{align}
The expansion $Z^{(CFT)}$ was obtained in \cite{MaMoAGT2,MaMoAGT3} via exact non-Gaussian contour integration over the contours $[0,q]$ and $[0,1]$ with multiplicities $N_1$ and $N_2$, just as indicated in (\ref{Dotsenko-Fateev}). This expansion coincides with standard series for the 4-point spherical conformal block, thus we call it a "CFT" expansion. To compare $Z^{(CFT)}$ with $Z^{(DV)}$ of the previous section, we need to compute the asymptotical $\hbar$-expansion both for the Gamma-functions in the normalisation constants $C_{N}(x,y)$ and for the rational expression for $B_1$. This computation is a straightforward exercise in undergraduate calculus and reduces essentially to use the Stirling formula

\begin{align}
\log \Gamma(x) = x \log x - x - \dfrac{1}{2} \log x + \dfrac{1}{2} \log(2 \pi) + \dfrac{1}{12 x} - \dfrac{1}{360 x^3} + O\left( \dfrac{1}{x^5} \right)
\end{align}
\smallskip\\
at large $x$. With dimensional analysis one can easily estimate that $Z^{(CFT)}$ has precisely the same form as $Z^{(DV)}$

\begin{align}
Z^{(CFT)} = {\widetilde C} \times \hbar^{{\widetilde \deg_{\hbar}}} \times q^{{\widetilde \deg_q}} \times \left( \sum\limits_{k = 0}^{\infty} \sum\limits_{m = -2k}^{\infty} {\widetilde Z_{km}} q^k \hbar^m \right)
\label{MainExpansion}
\end{align}
\smallskip\\
with some degrees ${\widetilde \deg_{\hbar}}, {\widetilde \deg_{q}}$ and coefficients ${\widetilde C}$, ${\widetilde Z_{km}}$, which need to be determined with the Stirling formula. Actual symbolic computation with the help of MAPLE shows, that ${\widetilde C} = C, \ {\widetilde \deg_{\hbar}} = \deg_{\hbar}, \ {\widetilde \deg_{q}} = \deg_{q}$ and

$${\widetilde Z_{km}} = Z_{km}, \ \ \ \ \mbox{ at least for } \ (k,m) = (0,0), (0,1), (0,2), (1,-2), (1,-1), (1,0)$$
\smallskip\\
thus suggesting that

\begin{align}
\boxed{
\ \ \ Z^{(CFT)} = Z^{(DV)} \ \ \
}
\label{Eq108}
\end{align}
\smallskip\\
This completes our check of the relation between two different methods to treat the 3-Penner ensemble and the two corresponding expansions. The detailed intermediate calculations with the Stirling formula, which lead to equalities ${\widetilde Z_{km}} = Z_{km}$ are too lengthy to be presented here. We only derive now the first three equalities, of normalisation constants ${\widetilde C} = C$ and of overall degrees ${\widetilde \deg_{\hbar}} = \deg_{\hbar}, \ {\widetilde \deg_{q}} = \deg_{q}$.

First of all, the overall degrees in $q$ coincide trivially: ${\widetilde \deg_{q}} = \deg_{q} = \deg$, so we are left only with the check of coincidence of overall degrees in $\hbar$ and of the normalisation constants. The small-$\hbar$ asymptotic Stirling expansion, necessary for such a check, includes terms of order $\hbar^{-1}, \hbar^{0}$ and $\log \hbar$:

\begin{align*}
\log C_{N_1}\left(\dfrac{\alpha_1}{\hbar}, \dfrac{\alpha_2}{\hbar}\right) + \log C_{N_2}\left(\dfrac{\alpha_1 + \alpha_2}{\hbar} + 2 \beta N_1, \dfrac{\alpha_3}{\hbar}\right) = \sum\limits_{a = 1}^{2} \log {\rm Vol}_\beta(N_a) + \emph{}
\end{align*}
\begin{align*}
\emph{} + \sum\limits_{k = 1}^{N_1} \left[ \log \Gamma\left(\dfrac{\alpha_1}{\hbar} + 1 + \beta(k-1)\right) + \log \Gamma\left(\dfrac{\alpha_2}{\hbar} + 1 + \beta(k-1)\right) - \log \Gamma\left(\dfrac{\alpha_1 + \alpha_2}{\hbar} + 2 + \beta(N_1+k-2)\right) \right] +
\end{align*}
\begin{align*}
\emph{} + \sum\limits_{k = 1}^{N_2} \left[ \log \Gamma\left(\dfrac{\alpha_1 + \alpha_2}{\hbar} + 2 \beta N_1 + 1 + \beta(k-1)\right) + \log \Gamma\left(\dfrac{\alpha_3}{\hbar} + 1 + \beta(k-1)\right) - \emph{} \right.
\end{align*}
\begin{align*}
\left. \emph{} - \log \Gamma\left(\dfrac{\alpha_1 + \alpha_2 + \alpha_3}{\hbar} + 2 \beta N_1 + 2 + \beta(N_2+k-2)\right) \right] =
\end{align*}
\begin{align}
\nonumber \ = \ & \sum\limits_{a = 1}^{2} \log {\rm Vol}_\beta(N_a) + \left( \dfrac{1 - \beta}{2} N_1 + \dfrac{\beta}{2} N_1^2 \right) \log ( \alpha_1 \alpha_2 \alpha_3 ) + \dfrac{1}{\hbar} N_1 (\alpha_1 \log \alpha_1 + \alpha_2 \log \alpha_2 + \alpha_3 \log \alpha_3) + \emph{} & \\ \nonumber & \\ \nonumber & \emph{} + \left( 2 \beta N_1 N_2 + \dfrac{3 - 3\beta}{2} N_1 + \dfrac{3 \beta}{2} N_1^2 + \dfrac{1 - \beta}{2} N_2 + \dfrac{\beta}{2} N_2^2 - \dfrac{1}{\hbar}(\alpha_1 + \alpha_2)(N_1 - N_2) \right) \log (\alpha_1 + \alpha_2) - \emph{} & \\ \nonumber & \\ \nonumber & \emph{} - \left( 2 \beta N_1 N_2 + \dfrac{3 - 3\beta}{2} N_2 + \dfrac{3 \beta}{2} N_2^2 + \dfrac{1}{\hbar}(\alpha_1 + \alpha_2 + \alpha_3) N_2 \right) \log (\alpha_1 + \alpha_2 + \alpha_3) + \emph{} & \\ \nonumber & \\ & \emph{} + \left( \dfrac{1 - \beta}{2} N_1 + \dfrac{1 - \beta}{2} N_2 + \dfrac{\beta}{2} N_1^2 + \dfrac{\beta}{2} N_2^2 \right) \log \hbar + O(\hbar^1) = \log C + (\deg_\hbar) \cdot \log \hbar + O(\hbar^1)
\end{align}
\smallskip\\
i.e, we indeed have ${\widetilde \deg_{\hbar}} = \deg_{\hbar}, \ {\widetilde \deg_{q}} = \deg_{q}$, as expected. Similarly, inclusion of higher order terms ($\hbar^1, \hbar^2$ and so on) in the Stirling expansion reproduces the coefficients $Z_{01}, Z_{02}$ and so on:

\begin{align}
\log C_{N_1}\left(\dfrac{\alpha_1}{\hbar}, \dfrac{\alpha_2}{\hbar}\right) C_{N_2}\left(\dfrac{\alpha_1 + \alpha_2}{\hbar} + 2 \beta N_1, \dfrac{\alpha_3}{\hbar}\right) = \log C + (\deg_\hbar) \cdot \log \hbar + Z_{01} \hbar + \left( Z_{02} - \dfrac{Z_{01}^2}{2} \right) \hbar^2 + \ldots
\end{align}
\smallskip\\
To reproduce the coefficients $Z_{km}$ with $k > 0$, it is essential to include the coefficients $B_k$ into consideration.

\pagebreak

\section{Conclusion}

In this paper we demonstrated the {\it consistency}
of two different treatments of the multi-Penner
$\beta$-ensemble, which plays an important role in
the study of AGT-relations.

\paragraph{$\bullet$}One side of equality is the original Dotsenko-Fateev
integral of \cite{MaMoAGT2,MaMoAGT3} with a system of
open integration contours taken with multiplicities
$(N_1,\ldots,N_r)$. In the case of $r=2$ the answer
was found in \cite{MaMoAGT2} and summarized in s.5
of the present paper, where it is called $Z^{(CFT)}$.

\paragraph{$\bullet$}Another side of equality is the same integral,
but treated differently: as Givental decomposition
\cite{DV, AMM} into a system of Gaussian models with
the filling numbers $(N_1,\ldots,N_r)$.
Above sections 3 and 4 are devoted to quasiclassical
evaluation of this decomposed integral,
in the form of a series in parameters $\alpha_1,\ldots,
\alpha_{r+1}$. The result for $r=2$ is exposed in
s.4, where it is called $Z^{(DV)}$.

\paragraph{}On the two sides of the equality the integrands
are the same, but integration contours are different.
Not-surprisingly, expansion coefficients in the two
cases {\it look} quite different, but in fact they
coincide(!), as stated in eq. (\ref{Eq108}).
A toy example of this phenomenon -- of contour
independence of quasiclassical expansion --
is analyzed in s.2. {\bf Significance} of the equality (\ref{Eq108}) between two different expansions

$$\boxed{Z^{(CFT)} = Z^{(DV)}}$$
\smallskip\\
is that to the r.h.s. one can apply the well developed
matrix-model technique.
In particular, $F = \log Z^{(DV)}$
should have a representation,
typical for the Seiberg-Witten (SW) theory \cite{SW}:
\be
\left\{ \begin{array}{c}
a_i = \oint_{A_i} \lambda, \\
\frac{\p F}{\p a_i} = \oint_{B_i}\lambda
\end{array}\right.
\ee
where the role of the Liouville-SW differential
is played by exact resolvent

$$\lambda = \ \left<\sum_i\frac{dz}{z-z_i}\right>$$
\smallskip\\
where the averaging is that of the multi-Penner $\beta$-ensemble. After this, the AGT relation implies that such generalized Seiberg-Witten prepotential is logarithm of
the $\epsilon_{1,2}$-regularized partition function
of ADHM super-instantons. This identification still remains to be understood.

\bigskip

In fact, (\ref{Eq108}) is not fully proved in this paper,
only the first terms of the $q,\hbar$-expansion
are explicitly evaluated and compared.
However, already this check is somewhat involved.

The difficult part is $Z^{(DV)}$.
Notably, $F = \log Z^{(DV)}$
is a well-known quantity.
It is actually the celebrated CIV-DV prepotential
\cite{DV}, just modified (generalized) in three
different directions:
from polynomial to logarithmic potential $V$,
to $\beta\neq 1$ and to arbitrary genus.

In fact, the {\it genus} expansion of matrix model
free energy is {\it not} the quasiclassical expansion of the sections 3 and 4.
Quasiclassical expansion is in $\hbar$ at given values of $N_a$, while
genus expansion is in $\hbar$ at given values of $S_a = \hbar N_a$.
These two expansions are of course related, but different.
We devote a big appendix below to demonstrate
that the same general DV-phase formulas (\ref{DV1}), (\ref{NormConst}), (\ref{DV2})) and (\ref{DV4})),
which are used to reproduce $Z^{(CFT)}$  for the
{\it logarithmic} potential $V$ in s.4, in another case
-- for the {\it polynomial} potential $V$ -- reproduce the well-known CIV-DV prepotential with above-mentioned modifications. This brings the well-developed theory of CIV-DV
prepotential into the circle of subjects,
embraced by the AGT relations.

\pagebreak

\section{Appendix. $\hbar$-expansion for the polynomial case. Generalized CIV-DV prepotential}

This appendix is devoted to another special case of general constructions of section 3 -- the special case of polynomial potentials $V(z)$ of arbitrary degree $\deg V(z) = r$:
\begin{align*}
V(z) = T_0 + T_1 z + \ldots + T_{r+1} z^{r+1}
\end{align*}
As in the generic case, these formulas are conveniently expressed not through the coefficients $T_a$, but rather through the critical points $\mu_a$ - roots of the derivative
\begin{align}
V^{\prime}(z) = (z - \mu_1) \ldots (z - \mu_r)
\end{align}
However, the polynomial case is certainly much simpler than the generic case, since all the quantities of interest -- including the values of potential and its derivatives at the critical points -- are completely determined here by the critical points theirselves. For example, for the 2nd derivatives we have
\begin{align*}
\Delta_a = V^{\prime \prime}(\mu_a) = \prod\limits_{a \neq b} (\mu_a - \mu_b)
\end{align*}
and expressions for higher derivatives are equally straightforward. These expressions need to be substituted into the generic quasiclassical formula for the partition function
\begin{align}
Z = Z_0 \times \prod\limits_{a = 1}^{r} {\rm Vol}_\beta(N_a) \times \left( 1 + \sum\limits_{k > 0} H_k \hbar^k \right)
\end{align}
or directly into the free energy
\begin{align}
F = \log Z = \log Z_0 + \sum\limits_{a = 1}^{r} \log {\rm Vol}_\beta(N_a) + H_1 \hbar + \left( H_2 - \dfrac{H_1^2}{2} \right) \hbar^2 + \left( H_3 - H_1 H_2 + \dfrac{H_1^3}{3} \right) \hbar^3 + \ldots
\end{align}
After the substitution, the answer becomes expressed entirely in terms of independent variables of the polynomial model: namely, the critical points $\mu_1, \ldots, \mu_r$, the filling numbers $N_1, \ldots, N_r$ and the parameter $\beta$. However, as explained in the Introduction, instead of the variables $N_a$ it is conventional to introduce different variables $S_a$ by the rule $S_a = \hbar N_a$. As follows from this rule, the number of integrations $N_a$ is no longer a constant: it becomes proportional to $\hbar^{-1}$. Quasiclassical expansion with such an additional requirement is usually called a genus expansion. Throughout this section we express the free energy in terms of variables $S_a$, thus dealing with a genus expansion, not just with quasiclassical expansion.

As it is easy to find out, the genus expansion of the free energy has a form
\begin{align}
F = \sum\limits_{k = 0}^{\infty} \ \Big[ \ G_{k/2}(S_1, \ldots, S_r) + F_{k/2}(S_1, \ldots, S_r) \ \Big] \ \hbar^{k-2}
\end{align}
with two types of contributions: the so-called perturbative prepotentials $G_{k/2}(S_1, \ldots, S_r)$ (they receive contributions from the normalisation constant $Z_0$ and the $\beta$-deformed group volume factor) and non-perturbative prepotentials $F_{k/2}(S_1, \ldots, S_r)$ (coming from the quantum part of the partition function, i.e, from the coefficients $H_k$). The lowest of these quantities -- $G_0(S_1, \ldots, S_r)$ and $F_0(S_1, \ldots, S_r)$ -- are known as perturbative and non-perturbative CIV-DV prepotentials, respectively.

\subsection{The case of $\beta = 1$ and genus zero: ordinary CIV-DV prepotential}

Using the above general formulas, one can easily find the perturbative \footnote{
Note, that the terms perturbative and non-perturbative refer to 4d gauge theory interpretation of these quantities, which does not coincide with their matrix-model meaning: in the matrix model, quantities $F_{k/2}(S_1, \ldots, S_r)$ come from the perturbative (in the small parameter $\hbar$) part $Z_{\hbar}$ of the partition function, while $G_{k/2}(S_1, \ldots, S_r)$ originate from the leading non-perturbative (in $\hbar$) part $Z_0$. This interesting phenomenon (perturbative gauge theory quantities correspond to non-perturbative matrix-model quantities and vice versa) can be considered as a concrete example of gauge-string duality.} part of the CIV-DV prepotential, by putting $\beta = 1$ and extracting the contributions of order $\hbar^{-2}$ (in terms of variables $S_a$) from the logarithms of both the normalisation constant and the group volume factor. The answer is
\begin{align}
G_{0}(S_1, \ldots, S_r)\Big|_{\beta = 1} = \sum\limits_{a = 1}^{r} \left( \dfrac{S_a^2}{2} \log \left( \dfrac{S_a}{\Delta_a} \right) - \dfrac{3}{4} S_a^2 - S_a V(\mu_a) \right)  - \sum\limits_{a \neq b} S_a S_b \log| \mu_a - \mu_b |
\end{align}
Similarly, the non-perturbative part of the CIV-DV prepotential can be found, extracting the contributions of order $\hbar^{-2}$ (in terms of variables $S_a$) from the logarithm of the quantum part $Z_{\hbar}$. The answer is an infinite series in $S$-variables. It is convenient to denote its homogeneous parts of degree $p$ as $F^{(p)}_{0}(S_1, \ldots, S_r)$:

\begin{align}
F_{0}(S_1, \ldots, S_r) = F^{(3)}_{0}(S_1, \ldots, S_r) + F^{(4)}_{0}(S_1, \ldots, S_r) + O(S^5)
\end{align}
\smallskip\\
For the cubic contribution, we find

\begin{align}
\nonumber F^{(3)}_{0}(S_1, \ldots, S_r)\Big|_{\beta = 1} \ = \ &  \sum\limits_{a \neq b} \dfrac{1}{\Delta_a (\mu_a - \mu_b)^2} \left( \dfrac{1}{2} S_a S_b^2  + \dfrac{3}{2} S_a^2 S_b + \dfrac{2}{3} S_a^3 \right) + \emph{} & \\ \nonumber & \\ & \emph{} + \sum\limits_{a \neq b \neq c} \dfrac{1}{\Delta_a (\mu_a - \mu_b)(\mu_a - \mu_c)} \left( \dfrac{1}{2} S_{a}^2 S_{b} + \dfrac{1}{2} S_{a} S_{b} S_{c} + \dfrac{1}{2} S_{a}^2 S_{c} + \dfrac{5}{12} S_{a}^3 \right)
\end{align}
\smallskip\\
For the quartic contribution, we find

\begin{align*}
F^{(4)}_{0}(S_1, \ldots, S_r)\Big|_{\beta = 1} \ = \ & \sum\limits_{a \neq b} \dfrac{1}{\Delta_a^2 (\mu_a - \mu_b)^4} \left( \dfrac{5}{6} S_{a} S_{b}^3+\dfrac{21}{4} S_{a}^2 S_{b}^2+\dfrac{47}{6} S_{a}^3 S_{b}+\dfrac{8}{3} S_{a}^4 \right) +
\end{align*}
\begin{align*}
+ \sum\limits_{a \neq b \neq c} \dfrac{1}{\Delta_a^2 (\mu_a - \mu_b)^3(\mu_a - \mu_c)} \left( 2 S_{a} S_{b}^2 S_{c}+\dfrac{1}{3} S_{a} S_{b}^3+7 S_{a}^2 S_{b} S_{c}+ \dfrac{9}{2} S_{a}^2 S_{b}^2+4 S_{a}^3 S_{c}+\dfrac{31}{3} S_{a}^3 S_{b}+\dfrac{14}{3} S_{a}^4 \right) +
\end{align*}
\begin{align*}
\nonumber & + \sum\limits_{a \neq b \neq c} \dfrac{1}{\Delta_a^2 (\mu_a - \mu_b)^2(\mu_a - \mu_c)^2} \left( \dfrac{3}{4} S_{a} S_{b} S_{c}^2+\dfrac{3}{4} S_{a} S_{b}^2 S_{c}+S_{a}^2 S_{c}^2 + \dfrac{19}{4} S_{a}^2 S_{b} S_{c}+S_{a}^2 S_{b}^2 + 4 S_{a}^3 S_{c} + \right. \emph{} & \\ \nonumber & \\ \nonumber & \left. \emph{} + 4 S_{a}^3 S_{b}+\dfrac{41}{16} S_{a}^4 \right) + \sum\limits_{a \neq b \neq c \neq d} \dfrac{1}{\Delta_a^2 (\mu_a - \mu_b)^2(\mu_a - \mu_c)(\mu_a - \mu_d)} \left( \dfrac{3}{2} S_{a} S_{b} S_{c} S_{d}+\dfrac{1}{2} S_{a} S_{b}^2 S_{d}+\dfrac{1}{2} S_{a} S_{b}^2 S_{c} + \right. \emph{} & \\ \nonumber & \\ \nonumber & \emph{} \left. + 2 S_{a}^2 S_{c} S_{d}+\dfrac{7}{2} S_{a}^2 S_{b} S_{d}+\dfrac{7}{2} S_{a}^2 S_{b} S_{c}+\dfrac{5}{4} S_{a}^2 S_{b}^2+3 S_{a}^3 S_{d}+3 S_{a}^3 S_{c}+\dfrac{11}{2} S_{a}^3 S_{b}+\dfrac{29}{8} S_{a}^4 \right) + \emph{}
\end{align*}
\begin{align}
\nonumber & \emph{} + \sum\limits_{a \neq b \neq c \neq d \neq e} \dfrac{1}{\Delta_a^2 (\mu_a - \mu_b)(\mu_a - \mu_c)(\mu_a - \mu_d)(\mu_a - \mu_e)} \left( \dfrac{1}{12} S_{a} S_{c} S_{d} S_{e} + \dfrac{1}{12} S_{a} S_{b} S_{d} S_{e} + \dfrac{1}{12} S_{a} S_{b} S_{c} S_{e} + \right. \emph{} & \\ \nonumber & \\ \nonumber & \left. \emph{} + \dfrac{1}{12} S_{a} S_{b} S_{c} S_{d}+\dfrac{5}{24} S_{a}^2 S_{d} S_{e}+\dfrac{5}{24} S_{a}^2 S_{c} S_{e}+\dfrac{5}{24} S_{a}^2 S_{c} S_{d}+\dfrac{5}{24} S_{a}^2 S_{b} S_{e}+ \dfrac{5}{24} S_{a}^2 S_{b} S_{d}+\dfrac{5}{24} S_{a}^2 S_{b} S_{c}+ \right. \emph{} & \\ \nonumber & \\ \nonumber & \emph{} \left. +\dfrac{1}{3} S_{a}^3 S_{e}+\dfrac{1}{3} S_{a}^3 S_{d}+\dfrac{1}{3} S_{a}^3 S_{c}+\dfrac{1}{3} S_{a}^3 S_{b}+\dfrac{119}{288} S_{a}^4
\right) + \emph{}
\end{align}

{\fontsize{8pt}{0pt}{
\be
\emph{} + \sum\limits_{a \neq b} \dfrac{1}{\Delta_a\Delta_b} \left( \dfrac{S_aS_b(4S_a^2 + 11S_a S_b + 4S_b^2) }{4(\mu_a - \mu_b)^4} - \mathop{\sum\limits_{c \neq a}}_{d \neq b} \dfrac{S_a S_b(S_a + S_c)(S_b + S_d)}{2 (\mu_a - \mu_b)^2 (\mu_a - \mu_c) (\mu_b - \mu_d)}
- \sum\limits_{c \neq b} \dfrac{S_aS_b(S_a+2S_b)(S_b+S_c)}{ (\mu_a - \mu_b)^3 (\mu_b - \mu_c)} \right)
\ee
}}
These formulas reproduce and generalise various explicit results, available in the literature \cite{DV}, in particular the cubic contribution coincides with the one obtained in \cite{DVItoyamaMorozov} by different methods.

\subsection{Generalisation to $\beta = 1$ and higher genera }

To generalise the CIV-DV prepotential to higher genera (i.e, to higher $k > 0$) one just needs to extract the contributions of higher orders $\hbar^{k-2}$ (in terms of variables $S_a$). Doing so with the normalisation constant and the group volume factor (actually, for $\beta = 1$ and $k > 0$ only the group volume factor contributes), one obtains
\begin{align}
G_{1/2}(S_1, \ldots, S_r)\Big|_{\beta = 1} = \sum\limits_{a = 1}^{r} S_a \left[ \log( 2 \pi ) - 1 + \log\left(\dfrac{S_a}{\hbar}\right) \right]
\end{align}
\begin{align}
G_{1}(S_1, \ldots, S_r)\Big|_{\beta = 1} = \sum\limits_{a = 1}^{r} \dfrac{5}{12} \log\left( \dfrac{S_a}{\hbar} \right) + \zeta^{\prime}(-1) + \dfrac{1}{2} \log(2\pi)
\end{align}
and, for all $m \geq 1$
\begin{align}
G_{m+1/2}(S_1, \ldots, S_r)\Big|_{\beta = 1} = \dfrac{B_{2m+2}}{(2m+1)(2m+2)} \sum\limits_{a = 1}^{r} \left( \dfrac{1}{S_a} \right)^{2m-1}
\end{align}
\begin{align}
G_{m+1}(S_1, \ldots, S_r)\Big|_{\beta = 1} = \dfrac{B_{2m+4}}{(2m+2)(2m+4)} \sum\limits_{a = 1}^{r} \left( \dfrac{1}{S_a} \right)^{2m}
\end{align}
\smallskip\\
where $B_{k}$ are the Bernoulli numbers:

$$ B_2=\frac{1}{6},\ B_{4} = -\frac{1}{30},\ B_{6} = \frac{1}{42}, \ldots, \ \ \mbox{with generating function} \ \ \ \sum_{k=2}^\infty \frac{B_{k} z^k}{k!} = \frac{z}{e^z - 1} - 1 + \dfrac{z}{2}$$
\smallskip\\
and $\zeta(x) = \sum_{i = 1}^{\infty} i^{-x}$ is the Riemann zeta function. Doing so for the quantum part $Z_{\hbar}$ of the partition function, one obtains non-perturbative prepotentials as infinite series in $S$-variables:

\begin{align}
F_{k/2}(S_1, \ldots, S_r) = \sum\limits_{p} F^{(p + 3-k)}_{k}(S_1, \ldots, S_r)
\end{align}
\smallskip\\
where $F^{(p)}_{k}(S_1, \ldots, S_r)$ denote homogeneous contributions of degree $p$. Notably, for $\beta = 1$ all the contributions of half-integer genus (corresponding to odd values of $k = 1,3,5,\ldots$) vanish: as we will see in the forthcoming section, they are proportional to $\beta - 1$. All the integer genera do contribute. For genus one, we obtain

\begin{align}
F_{1}(S_1, \ldots, S_r) = F^{(1)}_{1}(S_1, \ldots, S_r) + F^{(2)}_{1}(S_1, \ldots, S_r) + O(S^3)
\end{align}
\smallskip\\
where
\begin{align}
F^{(1)}_{1}(S_1, \ldots, S_r)\Big|_{\beta = 1} \ = \ & - \sum\limits_{a \neq b} \dfrac{S_a}{6\Delta_a (\mu_a - \mu_b)^2} + \sum\limits_{a \neq b \neq c} \dfrac{S_a}{24 \Delta_a (\mu_a - \mu_b)(\mu_a - \mu_c)}
\end{align}
\begin{align*}
\nonumber F^{(2)}_{1}(S_1, \ldots, S_r)\Big|_{\beta = 1} \ = \ & - \dfrac{1}{2} \sum\limits_{a \neq b} \dfrac{S_a S_b}{\Delta_a \Delta_b (\mu_a - \mu_b)^4} + \sum\limits_{a \neq b} \dfrac{1}{\Delta_a^2 (\mu_a - \mu_b)^4} \left( \dfrac{25}{12} S_{a} S_{b}+\dfrac{7}{3} S_{a}^2 \right)
+ \emph{} & \\ \nonumber & \\ & \emph{} +
\sum\limits_{a \neq b \neq c} \dfrac{1}{\Delta_a^2 (\mu_a - \mu_b)^3 (\mu_a - \mu_c)} \left( S_{a} S_{c}+\dfrac{11}{6} S_{a} S_{b}+\dfrac{17}{6} S_{a}^2 \right)
+ \emph{} & \\ \nonumber & \\ & \emph{} +
\sum\limits_{a \neq b \neq c} \dfrac{1}{\Delta_a^2 (\mu_a - \mu_b)^2 (\mu_a - \mu_c)^2} \left( \dfrac{3}{4} S_{a} S_{c}+\dfrac{3}{4} S_{a} S_{b}+\dfrac{23}{16} S_{a}^2 \right)
+ \emph{} & \\ \nonumber & \\ & \emph{} + \sum\limits_{a \neq b \neq c \neq d} \dfrac{1}{\Delta_a^2 (\mu_a - \mu_b)^2(\mu_a - \mu_c)(\mu_a - \mu_d)} \left( \dfrac{1}{2} S_{a} S_{d}+\dfrac{1}{2} S_{a} S_{c}+\dfrac{3}{4} S_{a} S_{b}+\dfrac{13}{8} S_{a}^2 \right) +
\end{align*}
{\fontsize{9pt}{0pt}{
\begin{align}
+ \sum\limits_{a \neq b \neq c \neq d \neq e} \dfrac{1}{\Delta_a^2 (\mu_a - \mu_b)(\mu_a - \mu_c)(\mu_a - \mu_d)(\mu_a - \mu_e)} \left( \dfrac{1}{24} S_{a} S_{e}+\dfrac{1}{24} S_{a} S_{d} + \dfrac{1}{24} S_{a} S_{c}+\dfrac{1}{24} S_{a} S_{b}+\dfrac{43}{288} S_{a}^2 \right)
\end{align}}}
To obtain explicit formulas for genus two and higher, it is necessary to calculate higher coefficients $H_k$ of the quantum partition function $Z_\hbar$. The above results are obtained, making use of coefficients $H_1$ and $H_2$ only. Higher $H_k$ can be straightforwardly calculated with the methods of section 3.

\subsection{Generalisation to arbitrary $\beta$ and genus zero }

To generalise the CIV-DV prepotential to arbitrary values of $\beta$, one just needs to keep $\beta$ as a free unconstrained parameter in all the formulas. This slightly boosts the complexity of intermediate algebraic calculations, but does not conceptually affect neither the method of calculation, nor the structure of the answer. Extracting the contributions of order $\hbar^{-2}$ from the logarithms of both the normalisation constant and the group volume factor, we find the $\beta$-deformed perturbative CIV-DV prepotential
\begin{align}
G_{0}(S_1, \ldots, S_r) = \sum\limits_{a = 1}^{r} \left( \dfrac{\beta S_a^2}{2} \log \left( \dfrac{S_a}{\Delta_a} \right) - \dfrac{3 \beta}{4} S_a^2 - S_a V(\mu_a) \right) - \beta \sum\limits_{a \neq b} S_a S_b \log| \mu_a - \mu_b |
\end{align}
I.e., at the spherical level the $\beta$-deformed perturbative prepotential depends on $\beta$ just linearly.
Similarly, the non-perturbative part of the CIV-DV prepotential is found extracting the contributions of order $\hbar^{-2}$ from the logarithm of the quantum part $Z_{\hbar}$. Direct calculation shows that
\begin{align}
F^{(3)}_{0}(S_1, \ldots, S_r) = \beta^2 \cdot F^{(3)}_{0}(S_1, \ldots, S_r)\Big|_{\beta = 1}
\end{align}
\begin{align}
F^{(4)}_{0}(S_1, \ldots, S_r) = \beta^3 \cdot F^{(4)}_{0}(S_1, \ldots, S_r)\Big|_{\beta = 1}
\end{align}
It is clearly seen, that the perturbative $\beta$-deformed CIV-DV prepotential depends on $\beta$ homogeneously:
\begin{align}
F_{0}(\lambda S_1, \ldots, \lambda S_r; \beta) = \lambda F_{0}(S_1, \ldots, S_r; \lambda \beta)
\end{align}
Equivalently, this statement can be expressed in a form of an exact differential equation:
\begin{align}
\beta \dfrac{\partial F_0}{\partial \beta} + F_0 = \sum\limits_{a = 1}^{r} S_a \dfrac{\partial F_0}{\partial S_a}
\end{align}
As we will see, this property fails at higher genera, but possibly gets substituted by some more sophisticated and yet unidentified property, which relates the dependence on $\beta$ and the dependence on $S$-variables.

\subsection{Generalisation to arbitrary $\beta$ and higher genera }

As already mentioned above, to handle the higher genera (i.e, higher $k > 0$) one just needs to extract the contributions of higher orders $\hbar^{k-2}$ in terms of variables $S_a$. For generic values of $\beta$, contributions of odd values of $k$ (non-integer genera $k/2$) become non-vanishing. The simplest of those is contribution of genus $1/2$.
\paragraph{Genus $1/2$.}  For the perturbative prepotential of genus $1/2$ we get
\begin{align}
G_{1/2}(S_1, \ldots, S_r) = \sum\limits_{a = 1}^{r} S_a \left[ \dfrac{1}{2} \log\left( \dfrac{S_a}{\Delta_a} \right) + \dfrac{\beta}{2} \log\left( \dfrac{S_a \Delta_a}{\hbar^2} \right) + \log\left(\dfrac{2\pi}{\Gamma(\beta)}\right) - \dfrac{1 + \log(\beta) + \beta - \beta\log(\beta)}{2} \right]
\end{align}
For the non-perturbative prepotential of genus $1/2$, we obtain

\begin{align}
F_{1/2}(S_1, \ldots, S_r) = F^{(2)}_{1/2}(S_1, \ldots, S_r) + F^{(3)}_{1/2}(S_1, \ldots, S_r) + O(S^4)
\end{align}
\smallskip\\
where the quadratic contribution has a form

\begin{align}
\nonumber F^{(2)}_{1/2}(S_1, \ldots, S_r) \ = \ & - \sum\limits_{a \neq b} \dfrac{\beta(\beta-1)}{\Delta_a (\mu_a - \mu_b)^2} \left( \dfrac{3}{2} S_{a} S_{b}+\dfrac{3}{2} S_{a}^2 \right) - \emph{} & \\ \nonumber & \\ & \emph{} - \sum\limits_{a \neq b \neq c} \dfrac{\beta(\beta-1)}{\Delta_a (\mu_a - \mu_b)(\mu_a - \mu_c)} \left( \dfrac{1}{2} S_{a} S_{c}+\dfrac{1}{2} S_{a} S_{b}+\dfrac{7}{8} S_{a}^2 \right)
\end{align}
\smallskip\\
and the cubic contribution has a form

\begin{align*}
F^{(3)}_{1/2}(S_1, \ldots, S_r) \ = \ & - \sum\limits_{a \neq b} \dfrac{\beta^2(\beta-1)}{\Delta_a^2 (\mu_a - \mu_b)^4} \left( \dfrac{21}{4} S_{a} S_{b}^2+\dfrac{71}{4} S_{a}^2 S_{b}+\dfrac{59}{6} S_{a}^3 \right) -
\end{align*}
\begin{align*}
\nonumber & \emph{} - \sum\limits_{a \neq b \neq c} \dfrac{\beta^2(\beta-1)}{\Delta_a^2 (\mu_a - \mu_b)^3(\mu_a - \mu_c)} \left( 7 S_{a} S_{b} S_{c}+\dfrac{9}{2} S_{a} S_{b}^2+9 S_{a}^2 S_{c}+\dfrac{45}{2} S_{a}^2 S_{b}+\dfrac{49}{3} S_{a}^3 \right)
- \emph{} & \\ \nonumber & \\ \nonumber & \emph{} - \sum\limits_{a \neq b \neq c} \dfrac{\beta^2(\beta-1)}{\Delta_a^2 (\mu_a - \mu_b)^2(\mu_a - \mu_c)^2} \left( S_{a} S_{c}^2+\dfrac{19}{4} S_{a} S_{b} S_{c}+S_{a} S_{b}^2+\dfrac{35}{4} S_{a}^2 S_{c}+\dfrac{35}{4} S_{a}^2 S_{b}+\dfrac{71}{8} S_{a}^3 \right) - \emph{} & \\ \nonumber & \\ \nonumber & \emph{} - \sum\limits_{a \neq b \neq c \neq d} \dfrac{\beta^2(\beta-1)}{\Delta_a^2 (\mu_a - \mu_b)^2(\mu_a - \mu_c)(\mu_a - \mu_d)} \left( 2 S_{a} S_{c} S_{d}+\dfrac{7}{2} S_{a} S_{b} S_{d}+\dfrac{7}{2} S_{a} S_{b} S_{c}+\dfrac{5}{4} S_{a} S_{b}^2+\dfrac{13}{2} S_{a}^2 S_{d}+ \emph{} \right. & \\ \nonumber & \\ \nonumber & \left. \emph{} +\dfrac{13}{2} S_{a}^2 S_{c}+\dfrac{47}{4} S_{a}^2 S_{b}+\dfrac{49}{4} S_{a}^3 \right) - \sum\limits_{a \neq b \neq c \neq d \neq e} \dfrac{\beta^2(\beta-1)}{\Delta_a^2 (\mu_a - \mu_b)(\mu_a - \mu_c)(\mu_a - \mu_d)(\mu_a - \mu_e)} \left( \dfrac{5}{24} S_{a} S_{d} S_{e} + \emph{} \right.
\end{align*}
{\fontsize{8pt}{0pt}{
\begin{align*}
\left. \emph{} + \dfrac{5}{24} S_{a} S_{c} S_{e}+\dfrac{5}{24} S_{a} S_{c} S_{d}+\dfrac{5}{24} S_{a} S_{b} S_{e}+\dfrac{5}{24} S_{a} S_{b} S_{d}+\dfrac{5}{24} S_{a} S_{b} S_{c}+\dfrac{17}{24} S_{a}^2 S_{e}+\dfrac{17}{24} S_{a}^2 S_{d}  + \dfrac{17}{24} S_{a}^2 S_{c} +\dfrac{17}{24} S_{a}^2 S_{b}+\dfrac{197}{144} S_{a}^3
\right) -
\end{align*}
\be
\emph{} - \sum\limits_{a \neq b} \dfrac{\beta^2(\beta-1)}{\Delta_a\Delta_b} \left( \dfrac{13 S_aS_b(S_a + S_b) }{4(\mu_a - \mu_b)^4} - \mathop{\sum\limits_{c \neq a}}_{d \neq b} \dfrac{S_a S_b(S_a + S_b + S_c + S_d)}{2 (\mu_a - \mu_b)^2 (\mu_a - \mu_c) (\mu_b - \mu_d)}
- \sum\limits_{c \neq b} \dfrac{S_aS_b(2 S_a + 3 S_b + 2 S_c)}{(\mu_a - \mu_b)^3 (\mu_b - \mu_c)} \right)
\ee
}}
\smallskip\\
Obviously, a variant of the homogeneity property holds here as well:
\begin{align}
F_{1/2}(\lambda S_1, \ldots, \lambda S_r; \beta) = \dfrac{\lambda \beta - \lambda}{\lambda \beta - 1} F_{1/2}(S_1, \ldots, S_r; \lambda \beta)
\end{align}
\paragraph{Genus 1.}For the perturbative prepotential of genus $1$, we obtain
\begin{align}
G_{1}(S_1, \ldots, S_r) = \dfrac{1 + 3 \beta + \beta^2}{12 \beta} \sum\limits_{a = 1}^{r} \log\left( \dfrac{S_a}{\hbar} \right) + \dfrac{\log(2\pi)}{4} + \dfrac{\beta}{12} + \dfrac{\gamma}{12\beta} - \beta \zeta^{\prime}(-1) + \sum\limits_{i = 1}^{\infty} \dfrac{B_{2i+2} \zeta(2i+1)}{2i(2i+1)\beta^{2i+1}}
\end{align}
where $\gamma = 0.57721\ldots$ is the Euler constant (this is an asymptotic expansion only, its reformulation in terms of convergent series or integrals remains to be found). For the non-perturbative prepotential of genus $1$, we obtain

\begin{align}
F_{1}(S_1, \ldots, S_r) = F^{(1)}_{1}(S_1, \ldots, S_r) + F^{(2)}_{1}(S_1, \ldots, S_r) + O(S^3)
\end{align}
\smallskip\\
where the linear contribution has a form

\begin{align}
\nonumber F^{(1)}_{1}(S_1, \ldots, S_r) \ = \ & - \sum\limits_{a \neq b} \dfrac{S_a}{\Delta_a (\mu_a - \mu_b)^2} \left( \dfrac{5}{6}-\dfrac{3}{2}\beta +\dfrac{5}{6} \beta^2 \right) + \emph{} & \\ \nonumber & \\ \nonumber & \emph{} + \sum\limits_{a \neq b \neq c} \dfrac{S_a}{\Delta_a (\mu_a - \mu_b)(\mu_a - \mu_c)} \left( \dfrac{11}{24}-\dfrac{7}{8} \beta + \dfrac{11}{24} \beta^2 \right)
\end{align}
and the quadratic contribution has a form

\begin{align*}
F^{(2)}_{1}(S_1, \ldots, S_r) = \sum\limits_{a \neq b} \dfrac{1}{\Delta_a^2(\mu_a - \mu_b)^4} \left[ \left( \dfrac{119}{12} \beta - \dfrac{71}{4} \beta^2 + \dfrac{119}{12} \beta^3 \right) S_a S_b + \left( \dfrac{73}{6} \beta - 22 \beta^2 + \dfrac{73}{6} \beta^3 \right) S_a^2 \right] + \emph{}
\end{align*}
\begin{align*}
\nonumber & \emph{} +
\sum\limits_{a \neq b \neq c} \dfrac{1}{\Delta_a^2(\mu_a - \mu_b)^3(\mu_a - \mu_c)} \left[ \left(5 \beta-9 \beta^2+5 \beta^3 \right)S_aS_c + \left(\dfrac{73}{6} \beta-\dfrac{45}{2} \beta^2+\dfrac{73}{6} \beta^3\right)S_aS_b+ \right.
 \emph{} \\ \nonumber & \\ \nonumber & \left. \emph{} +
\left(\dfrac{115}{6} \beta - \dfrac{71}{2} \beta^2 + \dfrac{115}{6} \beta^3\right)S_a^2 \right] +
\sum\limits_{a \neq b \neq c} \dfrac{1}{\Delta_a^2(\mu_a - \mu_b)^2(\mu_a - \mu_c)^2}
\left[ \left(\dfrac{19}{4} \beta-\dfrac{35}{4} \beta^2+\dfrac{19}{4} \beta^3\right) S_a (S_b + S_c) + \emph{}
\right.
 \emph{} \\ \nonumber & \\ \nonumber & \left. \emph{} + \left(\dfrac{165}{16} \beta-\dfrac{307}{16} \beta^2+\dfrac{165}{16} \beta^3\right) S_a^2
\right] +  \sum\limits_{a \neq b \neq c \neq d} \dfrac{1}{\Delta_a^2(\mu_a - \mu_b)^2(\mu_a - \mu_c)(\mu_a - \mu_d)} \times
\emph{} \\ \nonumber & \\ \nonumber & \emph{}
\times
\left[ \left(\dfrac{7}{2} \beta-\dfrac{13}{2} \beta^2+\dfrac{7}{2} \beta^3\right) S_a (S_c + S_d)
+ \left(\dfrac{25}{4} \beta - \dfrac{47}{4} \beta^2+\dfrac{25}{4} \beta^3\right) S_a S_b + \left(\dfrac{111}{8} \beta - \dfrac{209}{8} \beta^2 + \dfrac{111}{8} \beta^3\right) S_a^2 \right]
+ \emph{} \\ \nonumber & \\ \nonumber & \emph{} +
\sum\limits_{a \neq b \neq c \neq d \neq e} \dfrac{1}{\Delta_a^2(\mu_a - \mu_b)(\mu_a - \mu_c)(\mu_a - \mu_d)(\mu_a - \mu_e)} \left[ \left( \dfrac{3}{8} \beta - \dfrac{17}{24} \beta^2 + \dfrac{3}{8} \beta^3 \right) S_a (S_b + S_c + S_d + S_e) + \emph{} \right. \\ \nonumber & \\ \nonumber & \left. \emph{} +
\left( \dfrac{437}{288} \beta - \dfrac{277}{96} \beta^2 + \dfrac{437}{288} \beta^3 \right) S_a^2 \right] + \emph{}
\end{align*}
\begin{align}
\emph{} + \sum\limits_{a \neq b} \dfrac{1}{\Delta_a\Delta_b} \left( \dfrac{(\beta^2 + 9\beta(\beta-1)^2) S_aS_b}{4(\mu_a - \mu_b)^4} - \mathop{\sum\limits_{c \neq a}}_{d \neq b} \dfrac{\beta(\beta-1)^2 S_aS_b}{2 (\mu_a - \mu_b)^2 (\mu_a - \mu_c) (\mu_b - \mu_d)}
- \sum\limits_{c \neq b} \dfrac{2\beta(\beta-1)^2 S_aS_b}{(\mu_a - \mu_b)^3 (\mu_b - \mu_c)} \right)
\end{align}
In this case, the prepotential does not seem to enjoy any homogeneity properties.

\paragraph{Genus 3/2 and higher.}For the perturbative prepotential of genus $3/2$, we obtain
\begin{align}
G_{3/2}(S_1, \ldots, S_r) = \dfrac{1}{24} + \dfrac{1}{24 \beta}
\end{align}
For the non-perturbative prepotential of genus $3/2$, we obtain

\begin{align}
F_{3/2}(S_1, \ldots, S_r) = F^{(1)}_{3/2}(S_1, \ldots, S_r) + O(S^2)
\end{align}
\smallskip\\
where the linear contribution has a form

\begin{align}
\nonumber F^{(1)}_{3/2}(S_1, \ldots, S_r) \ = \ & \left(5-\dfrac{73}{6} \beta+\dfrac{73}{6} \beta^2-5 \beta^3\right) \sum\limits_{a \neq b} \dfrac{S_a}{\Delta_a^2(\mu_a - \mu_b)^4} + \emph{} \\ \nonumber & \\ \nonumber & \emph{} + \left(\dfrac{15}{2}-\dfrac{115}{6} \beta+\dfrac{115}{6} \beta^2-\dfrac{15}{2} \beta^3\right) \sum\limits_{a \neq b \neq c} \dfrac{S_a}{\Delta_a^2(\mu_a - \mu_b)^3(\mu_a - \mu_c)} + \emph{} \\ \nonumber & \\ \nonumber & \emph{} +
\left(4 - \dfrac{165}{16} \beta + \dfrac{165}{16} \beta^2 - 4 \beta^3 \right) \sum\limits_{a \neq b \neq c} \dfrac{S_a}{\Delta_a^2(\mu_a - \mu_b)^2(\mu_a - \mu_c)^2} + \emph{} \\ \nonumber & \\ \nonumber & \emph{} +
\left(-\dfrac{21}{4} +\dfrac{111}{8} \beta -\dfrac{111}{8} \beta^2+\dfrac{21}{4} \beta^3 \right) \sum\limits_{a \neq b \neq c \neq d} \dfrac{S_a}{\Delta_a^2(\mu_a - \mu_b)^2(\mu_a - \mu_c)(\mu_a - \mu_d)}
+ \emph{} \\ \nonumber & \\ \nonumber & \emph{} +
\left(\dfrac{9}{16}-\dfrac{437}{288} \beta+\dfrac{437}{288} \beta^2-\dfrac{9}{16} \beta^3\right)
\sum\limits_{a \neq b \neq c \neq d \neq e} \dfrac{S_a}{\Delta_a^2(\mu_a - \mu_b)(\mu_a - \mu_c)(\mu_a - \mu_d)(\mu_a - \mu_e)}
\end{align}
\smallskip\\
To obtain explicit formulas for non-perturbative prepotentials of genus two and higher, it is necessary to calculate higher ($k > 2$) coefficients $H_k$. Much simpler are perturbative prepotentials -- they are obtained by expansion of the group volume factor and, hence, do not require knowledge of complicated expressions for $H_k$ at all. For the perturbative prepotentials of half-integer genera we find
\begin{align}
G_{m+1/2}(S_1, \ldots, S_r) = \left( \dfrac{1}{2} + \dfrac{1}{2\beta^{2m-1}} \right) \dfrac{B_{2m+2}}{(2m+1)(2m+2)} \sum\limits_{a = 1}^{r} \left( \dfrac{1}{S_a} \right)^{2m-1}
\end{align}
For the perturbative prepotentials of integer genera we find
\begin{align}
G_{m+1}(S_1, \ldots, S_r) = - \left( \sum\limits_{s = 0}^{m+1} B_{2m-2s} B_{2s} \dfrac{\Gamma(2m)}{\Gamma(2s+1)\Gamma(2m-2s+3)} \ \beta^{1-2s} \right) \sum\limits_{a = 1}^{r} \left( \dfrac{1}{S_a} \right)^{2m}
\end{align}
This completes our description of $\beta$-deformed higher-genera CIV-DV prepotentials. It is important to note, that all the non-perturbative prepotentials described above, posess one and the same structure:

\begin{align}
F^{(3-2g)}_{g} = \sum\limits_{a \neq b \neq c} \dfrac{H_{(1,1)}(S_a,S_b,S_c)}{\Delta_a (\mu_a - \mu_b)(\mu_a - \mu_c)} + \sum\limits_{a \neq b} \dfrac{H_{(2)}(S_a,S_b)}{\Delta_a (\mu_a - \mu_b)^2}
\end{align}
\begin{align}
\nonumber F^{(4-2g)}_{g} \ = \ & \sum\limits_{a \neq b \neq c \neq d \neq e} \dfrac{H_{(1,1,1,1)}(S_a,S_b,S_c,S_d,S_e)}{\Delta_a^2 (\mu_a - \mu_b)(\mu_a - \mu_c)(\mu_a - \mu_d)(\mu_a - \mu_e)} +  \sum\limits_{a \neq b \neq c \neq d} \dfrac{H_{(2,1,1)}(S_a,S_b,S_c,S_d)}{\Delta_a^2 (\mu_a - \mu_b)^2(\mu_a - \mu_c)(\mu_a - \mu_d)} + \emph{} \\ \nonumber & \\ \nonumber & \emph{} + \sum\limits_{a \neq b \neq c} \dfrac{H_{(3,1)}(S_a,S_b,S_c)}{\Delta_a^2 (\mu_a - \mu_b)^3(\mu_a - \mu_c)}
+
\sum\limits_{a \neq b \neq c} \dfrac{H_{(2,2)}(S_a,S_b,S_c)}{\Delta_a^2 (\mu_a - \mu_b)^2(\mu_a - \mu_c)^2}
+
\sum\limits_{a \neq b} \dfrac{H_{(4)}(S_a,S_b)}{\Delta_a^2 (\mu_a - \mu_b)^4}
+ \emph{} \\ \nonumber & \\ \nonumber & \emph{} +
\sum\limits_{a \neq b} \sum\limits_{c \neq a, d \neq b} \dfrac{H_{(1,1),(1,1)}(S_a,S_b,S_c,S_d)}{\Delta_a \Delta_b (\mu_a - \mu_b)^2 (\mu_a - \mu_c) (\mu_b - \mu_d)} +
\emph{} \\ \nonumber & \\ & \emph{}
+
\sum\limits_{a \neq b} \sum\limits_{c \neq b} \dfrac{H_{(2),(1,1)}(S_a,S_b,S_c)}{\Delta_a \Delta_b (\mu_a - \mu_b)^3 (\mu_b - \mu_c)}
+
\sum\limits_{a \neq b} \dfrac{H_{(2),(2)}(S_a,S_b)}{\Delta_a \Delta_b (\mu_a - \mu_b)^4}
\end{align}
where $H_{(1,1)}, H_{(2)}$ are certain polynomials of degrees $3-2g$ in $S$-variables, while $H_{(1,1,1,1)}$, $H_{(2,1,1)}$, $H_{(3,1)}$, $H_{(2,2)}$, $H_{(4)}$, $H_{(1,1),(1,1)}$, $H_{(2),(1,1)}$, $H_{(2),(2)}$ are certain polynomials of degrees $4-2g$ in $S$-variables. Already the examples described above (for genera $g = 0,1/2,1,3/2$) provide enough evidence to conjecture, that different structures appearing in the generalized CIV-DV prepotentials are labeled by (collections of) Young diagrams. It would be interesting to check this conjecture and understand, what kind of structures actually appear in the next (third) order of $S$-expansion. Such understanding could be essential for future applications of the CIV-DV prepotentials as yet another special functions of string theory \cite{AMM}.

\section*{Acknowledgements}

Our work is partly supported by Russian Federal Nuclear Energy Agency, Federal Agency for Science and
Innovations of Russian Federation under contract 02.740.11.5194, by RFBR grants 10-01-00536, by joint grants 09-02-90493-Ukr, 09-02-93105-CNRSL, 09-01-92440-CE, 09-02-91005-ANF, 10-02-92109-Yaf-a, by CNRS (A.M.) and by Dynasty Foundation (Sh.Sh.).

\end{document}